\documentclass[useAMS,usenatbib,fleqn]{mnras}

\usepackage{tikz}
\usetikzlibrary{positioning}

\usepackage{import}

\usepackage{natbib}
\usepackage{subfig}
\usepackage{graphicx}
\usepackage{etex}
\usepackage{amsmath}
\usepackage{relsize}
\usepackage{amssymb}
\usepackage{hyperref}
\usepackage {float}
\usepackage{xfrac}
\usepackage[]{algorithm2e}
\usepackage{lmodern,adjustbox,booktabs,multirow}
\usepackage[english]{babel}
\usepackage[]{appendix}

\title[Detecting outliers in astronomical images with WGANs]{Detecting outliers in astronomical images with deep generative networks.}
\author[Margalef-Bentabol et al.]{Berta Margalef-Bentabol$^{1,2}$\thanks{Berta.Margalef@nottingham.ac.uk}, Marc Huertas-Company$^{{2},{3},{4},{5}}$, Tom Charnock$^{6}$,\newauthor  Carla Margalef-Bentabol$^{7}$, Mariangela Bernardi$^{1}$, Yohan Dubois$^{6}$,\newauthor Kate Storey-Fisher$^{8}$, Lorenzo Zanisi$^{9}$\\
$^{1}$Department of Physics and Astronomy, University of Pennsylvania, Philadelphia, PA 19104, USA\\
$^{2}$LERMA, Observatoire de Paris, PSL Research University, CNRS, Sorbonne Universit\'es, UPMC Univ. Paris 06,F-75014 Paris, France\\
$^{3}$Univerist\'e de Paris, 5 Rue Thomas Mann - 75013, Paris, France\\
$^{4}$Departamento de Astrof\'isica, Universidad de La Laguna, E-38206 La Laguna, Tenerife, Spain\\
$^{5}$Instituto de Astrof\'isica de Canarias, E-38200 La Laguna, Tenerife, Spain\\
$^{6}$Institut d'Astrophysique de Paris, UMR 7095, CNRS, UPMC Univ. Paris VI, 98 bis boulevard Arago, 75014 Paris, France\\
$^{7}$Babbly co, Toronto, ON, Canada\\
$^{8}$The Center for Cosmology and Particle Physics,New York University, New York, NY 10003, USA\\
$^{9}$Department of Physics and Astronomy, University of Southampton, Highfield SO17 1BJ, UK
}

\begin{document}

\maketitle

\begin{abstract}
With the advent of future big-data surveys, automated tools for unsupervised discovery are becoming ever more necessary. In this work, we explore the ability of deep generative networks for detecting outliers in astronomical imaging datasets. The main advantage of such generative models is that they are able to learn complex representations directly from the pixel space. Therefore, these methods enable us to look for subtle morphological deviations which are typically missed by more traditional moment-based approaches. We use a generative model to learn a representation of \emph{expected} data defined by the training set and then look for deviations from the learned representation by looking for the best reconstruction of a given object. In this first proof-of-concept work, we apply our method to two different test cases. We first show that from a set of simulated galaxies, we are able to detect $\sim90\%$ of merging galaxies if we train our network only with a sample of isolated ones. We then explore how the presented approach can be used to compare observations and hydrodynamic simulations by identifying observed galaxies not well represented in the models. 
\end{abstract}

\section{Introduction}

In recent years, the amount of astronomical data produced both by observations and simulations has exponentially increased in volume and complexity. This trend is expected to continue in the near future with surveys such as LSST and EUCLID becoming available. Processing and extracting meaningful physical information from these new datasets is a new challenge for the community.

To provide the necessary computational relief, machine learning techniques are becoming more and more popular as a way to address the increasing complexity. In particular, supervised machine learning has proven to be very successful when large amounts of labeled or annotated data are available for classification \citep[e.g][]{Huertas15, Edward17,Cabrera17, Jacobs17,Sreejith18, Dominguez18, Huertas18, Davidzon19}, regression \citep[e.g.][]{Tuccillo18, Pasquet19,Bonjean19} and segmentation \citep[e.g.][]{Boucaud19}. 

Supervised algorithms rely on annotated datasets and are thus, not well suited to the discovery of new types of unknown objects which will certainly be present in future surveys. In order to fully unlock the discovery potential of machine learning we have to leverage unlabeled data. Unsupervised algorithms aim to learn the underlying distribution of the data and find patterns without relying on annotated data. Such unsupervised methods can be, hence, used to detect objects whose properties deviate from the \emph{expected} or \emph{normal} objects given a data distribution. These abnormal objects are commonly referred to as outliers or anomalies. Anomaly detection is an active field in machine learning and has a broad range of applications ranging from fraud detection to surveilance \citep[e.g.][]{Zhang18, Frery17} to early diagnosis of disease outbreaks \citep{Wong03}.

In astronomy, outliers can represent artifacts in the data, pipeline errors or new physics. In the case of data or pipeline artifacts, it is important to further analyse them to better reduce systematics. On the other hand, any novel findings can potentially lead to interesting new science or objects which differ between models and observations. Several machine learning methods have already been successfully applied to detect outliers in astrophysical datasets, such as unknown classes of objects or objects belonging to rare classes. For example, self-organizing maps have been used to detect unusual quasars \citep{Fustes13} and spectroscopic outliers \citep{Meusinger12}, random forests have been used to detect anomalous SDSS spectra \citep{Baron17}, one-class support vector machines have been employed for novel detection in the WISE survey \citep{Solarz17} and clustering algorithms have been utilised to detect anomalous data in light-curves \citep{Protopapas06,Giles18}.

Anomaly detection in astronomical images is, however, a more complex task because of the high dimensionality (i.e. high number of parameters, in this case the number of pixels in the image) and limited amount of data. Traditional machine learning approaches, such as those mentioned earlier, typically rely on some reduced set of summary statistics (such as photometry, spectroscopic features or shape measurements), which usually discard a large amount of information about the image complexity and, therefore, can miss some subtle morphological anomalies. Deep generative models, on the other hand, are a modern class of unsupervised methods with the ability to learn complex representations of high-dimensional data in such a way that they can generate new examples drawn approximately from the same distribution as the original dataset. Generative Adversarial Networks \citep[GANs][]{Goodfellow14} provide a framework for training such deep generative models. They have gained popularity in recent years for their ability to produce extremely realistic images of everyday scenes \citep{Karras17,Radford15} and have also been used in astronomy to generate realistic galaxy images \citep{Ravanbakhsh16}. Recently, work has shown that GANs can be efficiently used for anomaly detection in medical image datasets \citep{Schlegl17, Murphy18}. They present a promising application for anomaly detection in astronomical images. One of the key advantages of a generative model-based approach for anomaly detection is that the model can learn to represent complex data directly from the pixel distribution without relying on specific galaxy properties, that could, otherwise, introduce biases from the methods used to obtain such properties.

This work, as the first in a series of studies, is to perform the first test of the ability of generative models to identify outliers in astronomical imaging datasets (i.e. images that are significantly different to the expected or `normal' data). We first test the approach in a well-defined sample of simulated galaxies from the cosmological hydrodynamic simulation Horizon-AGN \citep{Dubois14}. We define isolated galaxies as the normal objects and then quantify how frequently merging galaxies are detected as outliers. Secondly, we explore whether Wasserstein GANs \citep[WGANs, discussed in section~\ref{sec.wgan},][]{Arjovsky17, Gulrajani17} can be employed to quantify differences between observations and simulations. 

The paper is structured as follows: In Section \ref{sec.data} we present the data we use for this work. In Section \ref{sec.method} we explain the methodology for anomaly detection. Section \ref{sec.applications} is devoted to describing different possible applications. And finally, we summarize our findings in Section \ref{sec.summary}

\section{Data}\label{sec.data}
For this work, we use both simulated data from the Horizon-AGN cosmological hydrodynamical simulation \citep{Dubois14} and observed galaxies from the CANDELS survey \citep{Koekemoer11, Grogin11}.

\subsection{Horizon-AGN}
We refer the reader to \cite{Dubois14} for complete details of the simulation suite. Horizon-AGN is a cosmological hydrodynamical simulation run in a $L_\textrm{box} = 100 h^{-1} \textrm{Mpc}$ cube with initial conditions drawn from WMAP-7 cosmological parameters \citep{Komatsu11}. The total volume contains $1024^3$ dark matter (DM) particles, with a DM mass resolution of $M_\textrm{DM,res} = 8\times10^7 M_{\odot}$. The simulation is run with the adaptive mesh refinement code RAMSES \citep{Teyssier02}, and the initially uniform grid is refined down to a minimum cell size of $1\textrm{kpc}$ constant in physical length. Gas is allowed to cool down to $10^4\textrm{K}$ through H and He collisions with a contribution from metals using a \cite{Sutherland93} model. Gas is heated from a uniform UV background after $z_\textrm{reion}=10$ following \cite{Haardt96}. Star formation occurs in regions where the gas density reaches a critical density of $n_0=0.1\,\textrm{H}\,\textrm{cm}^{-3}$ and it is modelled with a Schmidt law: $\rho_{\ast}=\epsilon_{\ast}\rho/t_\textrm{ff}$, where $\rho_{\ast}$ is the star formation rate density, $\rho$ is the gas density, $\epsilon_{\ast}=0.02$ \citep[e.g.][]{Kennicutt98} the constant star formation efficiency and $t_\textrm{ff}$ is the local free-fall time of the gas. Stellar feedback is included assuming a \cite{Salpeter55} initial mass function (IMF), and occurs via stellar winds, supernovae type II and type Ia, with mass, energy and metal release of six chemical species: O, Fe, C, N, Mg and Si. Black hole (BH) feedback is also included in the simulation as modelled in \cite{Dubois12}, with BHs releasing energy in a quasar (heating) mode for a high accretion rate (Eddington ratio $> 0.01$) and in radio mode (jet) for low accretion rates (Eddington ratios $< 0.01$). 

In Horizon-AGN, galaxies are identified using the AdaptaHOP structure finder \citep{Aubert04} over the stellar distribution. The merger trees for the identified galaxies are built using the procedure outlined in \cite{Tweed09}, considering $758$ time steps that cover a redshift range spanning from $z = 7$ to $z = 0$ and with a time difference of $~ 17\ \textrm{Myrs}$ in average between two successive time steps.

\subsubsection{Mock images}
From the output of the simulation, we produce mock observations that will be used to train the generative models. In particular, mock images are produced to replicate the properties of the HST-CANDELS images in the H-band (F160W), using the SUNSET code \citep[e.g.][]{ Kaviraj17, Laigle19}, which models the emission of all galaxy particles to produce an image in the observed-frame. For each identified galaxy in the simulation, we define a cubic volume centered around the galaxy with an edge length of $8$ times the radius of the galaxy (in this case, defined as the average between the three semi-axes obtained when fitting an ellipsoid to the stellar mass distribution of the galaxy). This volume should contain the stellar particles from the main galaxy as well as those from any close companion, in order to capture any secondary progenitor in the image for the case of galaxy mergers. The stellar particles contained within the volume are used as an input to SUNSET, along with the spectral response of the H-band filter of the WFC3 camera. SUNSET computes the fluxes corresponding to the inputs using the stellar models of \cite{Bruzual03} and a \cite{Chabrier03} IMF. It is assumed that each particle is well described by a simple stellar population, for determining the contribution of each particle to the Spectral Energy Distribution (SED). For this work, we chose not to include dust effects in the image generation for computational reasons. This should not be a problem for tests involving only simulated data but can significantly affect the comparison with observations. We will discuss this in section \ref{subsec.candels}. Finally, the integration of the SED in each pixel and the redshift of the galaxy are used to generate an image in the observed frame. The physical size of the pixel is re-scaled for every image to $0.06$ arcsec, to match the resolution of the CANDELS H-band images. The flux is then scaled using the H-band zero-point of CANDELS to match the S/N. Finally, to generate a realistic mock observation, the re-scaled images are convolved with the corresponding PSF. These steps are repeated for three different projections along the main axis of the simulations (X,Y,Z) so that for every 3D cube three images are produced. The final sample built that way consists of $1,524,118$ mock images which include all galaxies with $\log (M_{\ast}/M_{\odot})>10$ and $0.5<z<3$, with $250$ snapshots in the redshift range. For the purpose of this work, we set a fixed image size of $64\times 64$ pixels.

\subsection{CANDELS}\label{sec.data.candels}

We use H-band images from the $5$ CANDELS \citep{Grogin11,Koekemoer11} fields: UDS, COSMOS, GOODS-S, GOODS-N and EGS. The parent sample comes from the catalogue of \cite{Dimauro18}. Our final selection is made of H-band selected galaxies with magnitudes brighter than $F160=23.5$, $\log (M_{\ast}/M_{\odot})>10$ and $0.5<z<3$, to match the sample of galaxies from the Horizon-AGN simulation in stellar mass and redshift. The final sample consists of $17,611$ CANDELS images.

We use the official catalogues of redshifts (spectroscopic redshifts are used when available) and stellar masses from CANDELS. The UDS and GOODS-S photometric redshifts were determined using the method described in \cite{Dahlen13}. Stellar masses are drawn from the catalogue presented in \cite{Santini14} using these photometric redshifts. For the COSMOS, GOODS-N and EGS fields, the photometric redshifts and stellar masses are discussed in \cite{Nayyeri17}, \cite{Barro19} and \cite{Stefanon17} respectively.

We additionally use the structural parameters (S\'ersic index, $n$, effective radius, $R_\textrm{eff}$ and axis ratio, $q$) published in \cite{Dimauro18}, obtained from 2D single S\'ersic fits on the H-band (F160W), and the deep-learning based visual morphologies from \cite{Huertas15}.

\section{Method}\label{sec.method}

\subsection{Deep neural networks}

Artificial Neural Networks \citep[ANN,][]{Hassoun95}, are computational techniques vaguely inspired by the connections that are established between the neurons in the brain and their ability to store and process information. An ANN consists of a collection of connected nodes or units (or neurons). The connections (or synapsis) are directed and have associated weights. Those weights are determined by training (or learning).

The nodes of a network are typically arranged in layers. The particular arrangement of the nodes into layers and the connection patterns between them is called the architecture of the neural network. Input layers contain the nodes that receive their input from an external source, output layers provide the output of the network and the layers in between are referred to as hidden layers. Each node of the hidden layers (hidden unit) is a mathematical function that receives inputs from units in the previous layer and computes an output that is transmitted to other units in the next layer based on the connecting weights. The goal is to use the network as a complex non-linear function that provides some desired output for each input. A cost function or loss is defined to quantify how far is the desired output from the network's actual output. Training is the process of determining the best set of weights to minimize the cost function for a given dataset.

Deep-learning Networks (or Deep Networks) are ANNs comprised of many more hidden layers than traditional ANNs and typically have more complex architectures and mathematical functions in their units. Convolutional Neural Networks (CNNs) are a particular architecture of deep networks that were developed within the context of image processing and computer vision applications \citep{Fukushima88}. CNNs are comprised of one or more convolutional layers followed by one or more fully-connected layers. The convolutional layers take as input a set of feature maps (e.g. the colour channels of an image) and convolve each of these with a set of learnable filters to produce the output feature maps. Each layer adds more abstraction to the original input and produces a more informative set of features for the next layer. The fully-connected layers have every node in a layer connected to every node in the following layer. They act as a classifier, taking as input the features from the last convolutional layer. The architecture of a CNN is designed to take advantage of the 2D structure of an input image, preserving the spatial relationship between pixels, i.e. they are able to learn translationally invariant features from the data. By exploiting the translational symmetry of the data, CNNs have shown to produce great results for pattern recognition in images.

\subsection{Generative Advesarial Networks, GANs}

Generative adversarial networks (GANs) were first introduced in \cite{Goodfellow14}. In the original formulation, they consist of two networks that are trained simultaneously: a generator (the generative model to be trained) and a discriminator (a classification model). The generator, $G_\theta$, with parameters $\boldsymbol{\theta}$, produces new samples from the approximated target data distribution whilst the discriminator, $D_\psi$, with parameters $\boldsymbol{\psi}$, aims to distinguish the generated samples from the true target distribution. 

The input for the generator is a random vector, $\boldsymbol{z}$, usually drawn from a normal or uniform distribution, and the output is drawn from the approximate target distribution $\widetilde{\boldsymbol{x}} = G_\theta(\boldsymbol{z})$, usually an image, although not necessarily so. The discriminator is trained as a standard classifier optimized to distinguish real and generated images. The output of the discriminator describes whether the features are likely to be from the true distribution or not. Once the discriminator is trained to optimise the parameters, $\psi$, for a given set of $\theta$, the discriminator parameters are fixed and the generator is trained to maximise the distance from the category designated as a \emph{generated} image. In doing so, the features which distinguish the two categories in the discriminator are backpropagated through to the generator, allowing the generator to create generated images with the features representative of the real ones. The networks keep training alternately until a Nash equilibrium \citep{Nash50} is reached and the generator creates images that are equally categorised by the discriminator as the real ones. The objective of the combined networks can be formulated as the minimax objective of distance $V(P_\textrm{r},P_\textrm{g})$ as follows:

\begin{align}
    V(P_\textrm{r},P_\textrm{g})=&\underset{\boldsymbol{\theta}\in\mathbb{R}^{N_G}}{\min}\ \underset{\boldsymbol{\psi}\in\mathbb{R}^{N_D}}{\max} \underset{\boldsymbol{x}\sim P_\textrm{r}}{\mathbb{E}}	[\log D_\psi(\boldsymbol{x})]\nonumber\\
    &\phantom{hello}+ \underset{\boldsymbol{z}\sim P_\textrm{z}} {\mathbb{E}}[\log(1-D_\psi(G_\theta(\boldsymbol{z})))]\,
\end{align}

\noindent where $P_\textrm{r}$ is the real data distribution and $P_\textrm{g}$ is the distribution of samples of generated targets $\widetilde{\boldsymbol{x}}=G_\theta(\boldsymbol{z})$ obtained from the latent distribution $\boldsymbol{z}\sim P_\textrm{z}$.

\subsection{Wasserstein generative adversarial networks, WGANs}\label{sec.wgan}

GANs have shown unprecedented achievements for many generative tasks, particularly in image generation. However, the original GAN formulation often suffers from convergence problems, when the network fails to find a Nash equilibrium \citep{Salimans16}, or mode collapse, that results in the generator producing limited varieties of samples. Since GANs were introduced, several improvements have been proposed in the literature that help stability in the training phase \citep[e.g.][]{Salimans16,Neyshabur17, Hoang19}. One of them is the Wasserstein-GAN (WGAN) model which is based on the Wasserstein-1 distance as the metric to measure the similarity between a real and a generated distribution. This type of network has been shown to be more stable and reach convergence more easily than the original formulation of GANs and prevents mode collapse \citep{Arjovsky17}. Although similar in style to traditional GANs, WGANs are theoretically separate. In principle, the difference with WGANs is that the discriminator, $D_\psi$ is replaced by another network, often called a critic, $C_\psi$. Instead of classifying images into real or generated categories, the critic gives an estimation of the Wassestein distance, which describes the amount of work necessary to transport a generated distribution to a target one. In our case, the distribution is the pixels of a collection of images. In times gone by, the Wasserstein distance has also been known as the Monge-Amp\`ere-Kantorovich distance and the Earth-mover distance (EMD). The name EMD arises since one can think of a probability distribution as a pile of earth where the EMD would be the minimal work needed to move one pile to the other. Work is defined as the amount of earth/mass that was moved times the travelled distance. Mathematically, the Wasserstein distance between two probability distributions $P_\textrm{r}$ and $P_\textrm{g}$ can be expressed as the supremum over the set of all 1-Lipschitz functions, $\boldsymbol{C}$, via:

\begin{equation}\label{eq.wasserstein}
    W(P_\textrm{r},P_\textrm{g})=\underset{C_\psi\in\boldsymbol{C}}{\sup}\left[  \underset{\boldsymbol{x}\sim P_\textrm{r}}{\mathbb{E}}[C_\psi(\boldsymbol{x})] \, - \underset{\boldsymbol{z}\sim P_\textrm{z}}{\mathbb{E}}[C_\psi(G_\theta(\boldsymbol{z}))] \,\right]_{\boldsymbol{\theta}=\boldsymbol{\theta}^*}
\end{equation}

\noindent where $\boldsymbol{\theta}^*$ is some fixed set of parameters of the generator. In order to implement a WGAN, we approximate the 1-Lipschitz functions in equation \eqref{eq.wasserstein} with a neural network, i.e. the critic, that is trained by maximizing the following cost function:

\begin{equation}
    L = \left[\underset{\boldsymbol{x}\sim P_\textrm{r}}{\mathbb{E}}[C_\psi(\boldsymbol{x})] - \underset{\boldsymbol{z}\sim P_\textrm{z}}{\mathbb{E}}[ \, C_\psi(G_\theta(\boldsymbol{z}))]\right]_{\boldsymbol{\theta}=\boldsymbol{\theta}^*}
\end{equation}\label{eq.single_loss}

However, the function $C_\psi$ learned by the critic has to be a 1-Lipschitz function in order to calculate the approximate Wasserstein distance. A differentiable function is 1-Lipschitz if and only if it has gradients with norm at most $1$ everywhere. This can be enforced in the WGAN using gradient penalty that penalizes the model if the gradient norm moves away from norm value of order unity, which results in adding a regularization term in the loss function. Therefore, the new loss function for the critic that we maximise has the following form:
\begin{align} 
    L_C =& \bigg[\underset{\boldsymbol{x}\sim P_\textrm{r}}{\mathbb{E}}[C_\psi(\boldsymbol{x})]
    -\underset{\boldsymbol{z}\sim P_z}{\mathbb{E}}[C_\psi(G_\theta(\boldsymbol{z})]\nonumber\\
    &\phantom{hellohello}+\lambda \underset{\hat{\boldsymbol{x}}\sim P_{\hat{\textrm{x}}}}{\mathbb{E}}[(\|\nabla_{\hat{\boldsymbol{x}}} C_\psi(\hat{\boldsymbol{x}})\|_2-1)^2 ]\bigg]_{\boldsymbol{\theta}=\boldsymbol{\theta}^*}
\end{align}\label{eq.loss}

\noindent where $\hat{\boldsymbol{x}}=\epsilon\boldsymbol{x}+(1-\epsilon)\widetilde{\boldsymbol{x}}$ is uniformly sampled from the straight line between a pair of data points sampled from the distribution of $P_r$ and samples of $\widetilde{\boldsymbol{x}}\equiv G_\theta(\boldsymbol{z})$ with $\boldsymbol{z}\sim P_z$.
$\epsilon$ is a mixing parameter, uniformly sampled between 0 and 1. $\lambda$ (the gradient penalty) is a hyperparameter that is, in practice, tuned to achieve optimal performance. 

Since the parameters $\boldsymbol{\theta}$ of the generator $G_\theta$ do not enter into the first term of equation \eqref{eq.single_loss}, its derivative with respect to $\boldsymbol{\theta}$ is zero and as such we can define the generator-only loss as:

\begin{equation}
    L_G =  \left[-\underset{\boldsymbol{z}\sim P_z}{\mathbb{E}}[ C_\psi(G_\theta(\boldsymbol{z}))]\right]_{\boldsymbol{\psi}=\boldsymbol{\psi}^*}
\end{equation}\label{eq.gen_loss}

\subsection{Training procedure}

\begin{figure}
    \centering
    \includegraphics[width=1\linewidth]{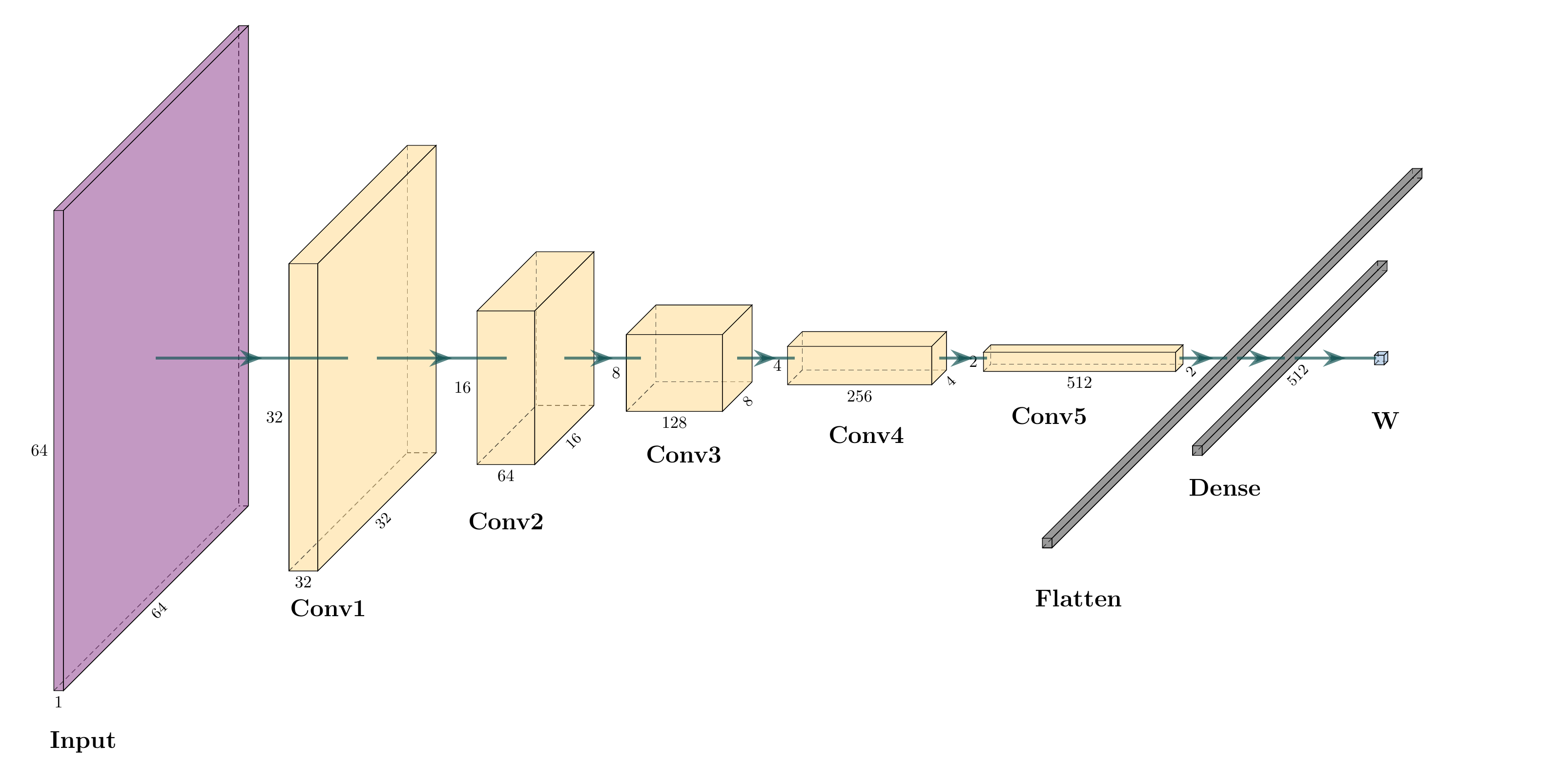}
    \includegraphics[width=1\linewidth]{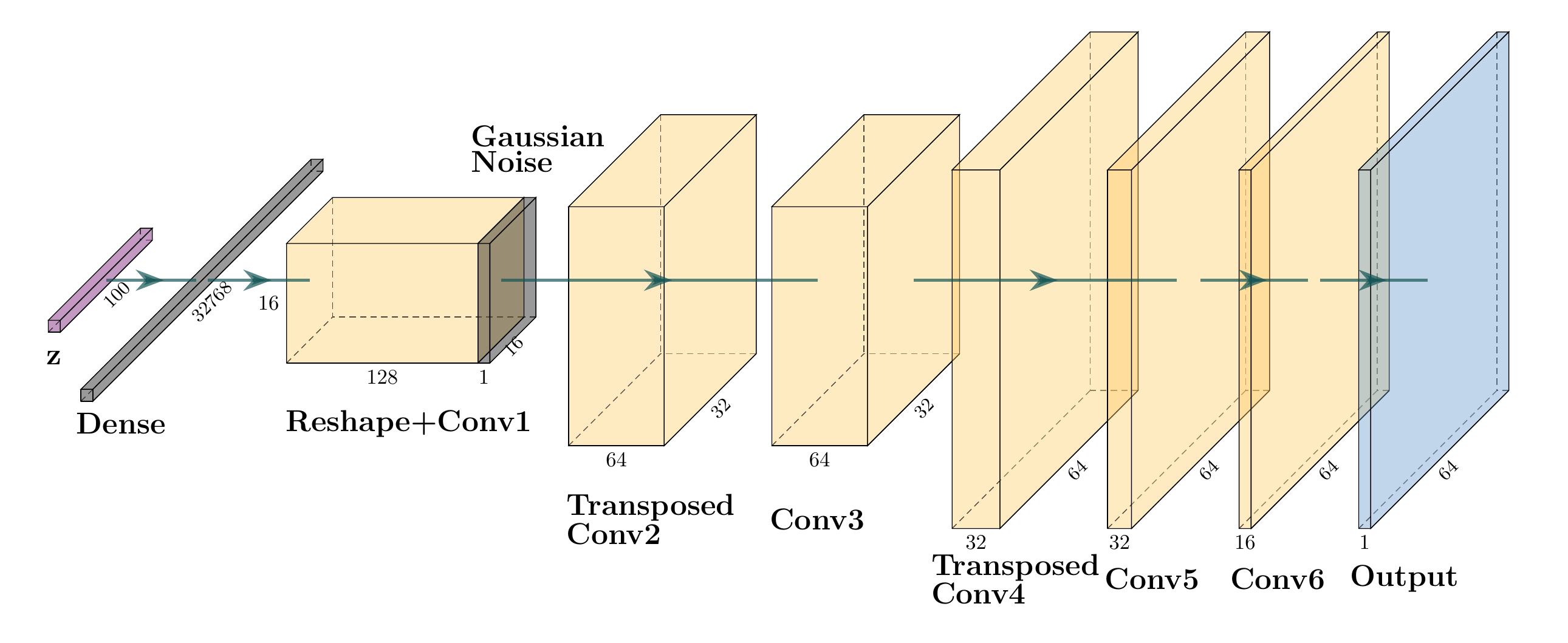}
    \caption{Critic (top) and generator network (bottom). The critic in this work consists of $5$ convolutional layers and $2$ dense layers. It takes as input an image of $64\times 64$ pixels and outputs a real number. The generator network is made of $1$ dense layer and $6$ convolutional layers. It takes as an input a random vector of size $100$ and outputs an image of size $64\times 64$ pixels.}\label{figure1}
\end{figure}

We implement CNN architectures for the critic and the generator, both shown in Figure \ref{figure1}. The generator network takes as input a random vector of size $100$ and outputs an image of size $64\times 64$ pixels. It consists of $1$ dense layer and $6$ convolutional layers. The critic consists of $5$ convolutional layers and $2$ dense layer. It takes as input an image of $64\times 64$ pixels and outputs a real number. The final architectures used in this work have been achieved through manual optimization and will not necessarily suit other applications. 

\begin{figure}
    \centering
    \includegraphics[width=1\linewidth]{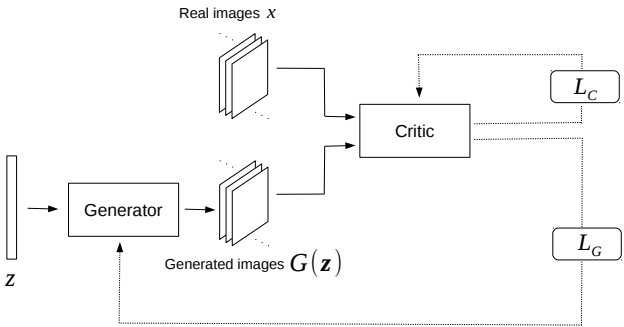}
    \caption{Schematic representation of the WGAN training. Given a batch of real and generated images, the critic is trained for $n_\textrm{critic}$ iterations to approximate the Wasserstein distance, by minimizing $L_c$ whilst keeping the weights of the generator fixed. Afterwards, the generator's weights are updated for a single iteration, whilst the critic weights are held constant so that it minimizes the approximate Wasserstein distance.}\label{figure2}
\end{figure}

The WGAN is trained following the standard procedure outlined in \cite{Gulrajani17}. Similarly to the original GAN, the two networks of the WGAN are trained alternately. A Schematic representation of the WGAN training is shown in Figure \ref{figure2}. We use a default value of $\lambda =10$ for the gradient penalty coefficient, a number of critic iterations per generator iteration $n_{critic}=10$, batch size $m=32$ and Adam optimizer with the following hyperparameters: $\alpha=0.00005$, $\beta_1=0.5$, $\beta_2=0.9$. We use Keras\footnote{https://keras.io/} with Tensorflow\footnote{https://www.tensorflow.org/} as the backend. The exact algorithm is shown in Appendix \ref{appendix}.

\subsection{Anomaly detection method}

Our goal is to use the trained WGAN to detect outliers. The main underlying idea is that after training is completed, the generator $G_\theta$ should be able to take a point $\boldsymbol{z}$ from the latent space and generate an image that resembles the images used for training (normal images). However, whenever an image does not come from the distribution of normal images then it will not be possible to generate a similar image from any point, $\{\boldsymbol{z}|\boldsymbol{z}\in\mathbb{R}^{N_z}\}$, in the latent space, $\mathbb{R}^{N_z}$ and it will be in some sense anomalous. 

Therefore, in order to identify if a given image $\boldsymbol{x}_t$ is an outlier, we need first to look for the closest image the trained network can generate from the latent space and then quantify the degree of similarity between the generated image $\widetilde{\boldsymbol{x}}'\equiv G_\theta(\boldsymbol{z}')$ and the original one $\boldsymbol{x}_t$. In this work, we follow the method described in \citep{Schlegl17} to find the $\boldsymbol{z}'$ vector that generates the closest image to a given input image. With the weights of the WGAN fixed, we train a neural network $\mu_\phi$ composed of $2$ fully connected layers that maps a noise vector, $\boldsymbol{y}$, of the same size as the latent space into the actual latent space, $\boldsymbol{z}$. This output is fed to the WGAN to generate an image. The shallow network is optimized using a loss function with two components, a residual loss $L_R$ and a critic loss $L_F$. The residual loss enforces the visual similarity pixel to pixel between the generated image $G_\theta(\boldsymbol{z})$ and $\boldsymbol{x}_t$. The critic loss pushes the generated image to lie on the learned manifold of trained images (i.e. have the same types of features). The total loss is defined as the weighted sum of both components:

\begin{equation}
    L_A=\gamma L_R+(1-\gamma) L_F
\end{equation}\label{eq.loss_anomaly}

\begin{figure}
    \centering
    \includegraphics[width=1\linewidth]{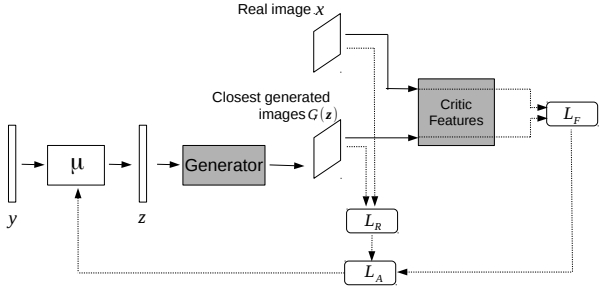}
    \caption{Schematic representation of the anomaly detection training. The grey color represents that the weights of a network are fixed during training. Given a real image and a noise vector, the network $\mu$ finds the anomaly score and the closest generated image by minimizing the combined loss $L_A$ (see equation \protect\eqref{eq.loss_anomaly}) whilst keeping the weights of the generator and critic fixed. The Critic features box represent the critic network without the last two dense layers, and it outputs a feature map obtained before the fully connected layers}\label{figure3}
\end{figure}

\noindent $\gamma \in(0,1)$ is a hyperparameter that weights the two contributions to the final loss. Here we use a value of $\gamma = 0.7$ (we note that our results do not change significantly when choosing different values of $\gamma$). Each contribution is defined as follow:

\begin{align} 
    L_R(\boldsymbol{z}') &=  |\boldsymbol{x} - G_\theta(\boldsymbol{z}')|\\
    L_F(\boldsymbol{z}') &=  |c_\varphi(\boldsymbol{x}) - c_\varphi(G_\theta(\boldsymbol{z}'))|
\end{align}

\noindent where $c_\varphi$ is the output of the last convolutional layer of the critic, $C_\psi$ , i.e. the set of informative features obtained before the fully connected layers. A Schematic representation of the training for the network $\mu_{\phi}$ is show in Figure \ref{figure3}.

An anomaly score AS is then defined as the loss at the last iteration, when the training has converged (i.e. the loss is not decreasing any further; in this case convergence is reached after about $500$ iterations) and the closest image in terms of equation~\eqref{eq.loss_anomaly} has been found:

\begin{equation}
    AS=\gamma L_R^o+(1-\gamma) L_F^o
\end{equation}

\noindent where $L_R^0$ and $L_F^0$ are the residual and critic loss at the last iteration, respectively.

The procedure is not optimal from a performance perspective since images need to be processed individually. It can easily be improved by performing a global optimization over multiple images and then applying a simple gradient descent to refine as shown in Storey-Fisher et al. (in prep). Since computing time is not critical for this work in which the sample of images to test is not enormous, we keep this original implementation.

Note that, whilst the anomaly score does not have a true meaning independent of the training of each of the critic, the generator and $\mu_\phi$, anomalous data can be identified by comparing it to the distribution of the training data, which the generator is trained to approximately draw from. For this reason, we measure the anomaly score for all the images in our training set which acts as the calibration of the anomaly detector. Any new image with an anomaly score significantly outside the bulk of scores for the training images is then quantified as being, in some way, anomalous.

\section{Applications}\label{sec.applications}

In the following section, we explore several cases in astronomy for our WGAN-based anomaly detector. We first calibrate how the anomaly detector performs with a sample of known anomalies. In particular, we quantify how accurately images of mergers are detected as anomalous when our training sample (the `normal' sample) is a set of isolated galaxies. We split this application in two cases: the first case focuses on detecting anomalies due to the presence of a neighbour object, and the second one focuses on analysing more subtle merger-induced morphological disturbances. The last application consists of using the anomaly score to compare a sample of images from the Horizon-AGN simulation to real galaxies from the CANDELS survey, to quantify the difference or similarity between such datasets. 

\subsection{Galaxy mergers as anomalies}

\subsubsection{Training}

\begin{figure*}
    \centering
    \includegraphics[trim={0 6cm 0 0},clip,width=1\linewidth]{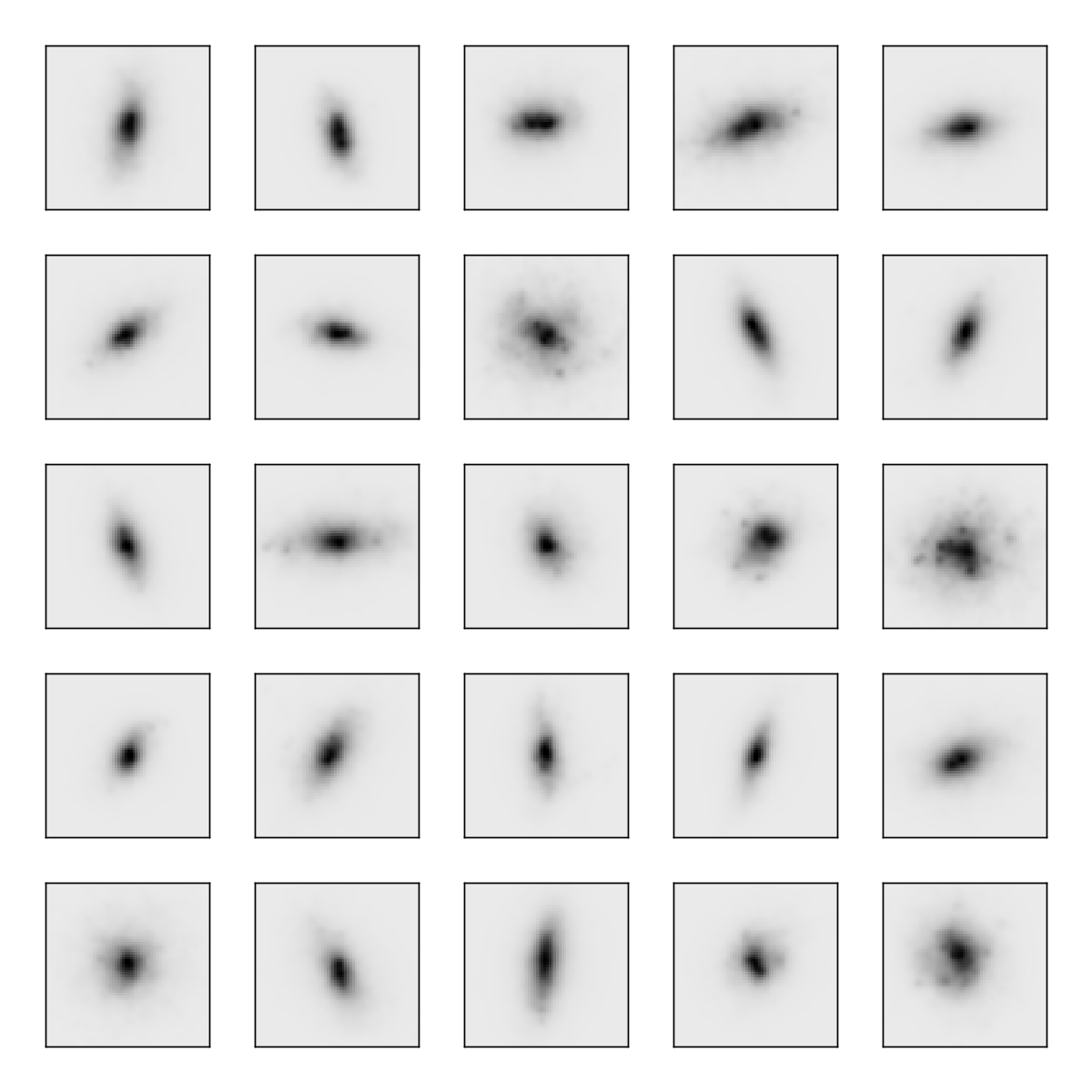}
    \caption{Examples of mock images (observed in the H-band) from the Horizon-AGN simulation that are used as training data. The images have a size of $64\times 64$ pixels, and the pixel scale is $0.06''/\textrm{pixel}.$ }\label{figure4}
\end{figure*}

The training set consists of images of isolated galaxies (no mergers), which we call the `normal images', and the test set consists of images of galaxy mergers. The selection of interacting galaxies in the simulations is done by checking an increase in galaxy mass due to the contribution of more than one progenitor from the previous time step \citep{Rodriguez-Gomez15, Abruzzo18}. If a galaxy has more than one progenitor and the ratio between the stellar mass provided by the secondary and the main progenitor is equal or larger than $1:10$, then that galaxy will be considered a merger. When a merger is identified, we build a \textit{merger sequence} by going back in time in the merger tree until the companion is four effective radii away from the central galaxy. We call all these images pre-mergers. We also follow forward in time after the merger event for the same number of time steps. These images are called post-mergers.

\begin{figure*}
    \centering
    \includegraphics[width=0.45\linewidth]{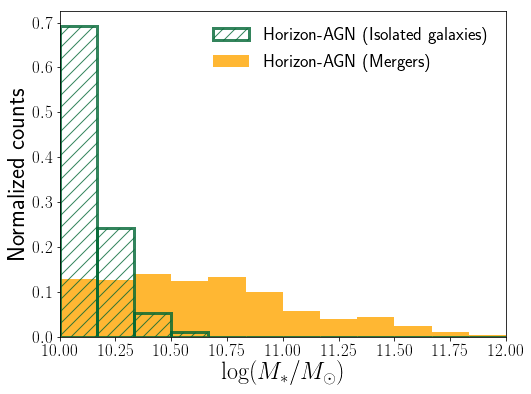}
    \includegraphics[width=0.45\linewidth]{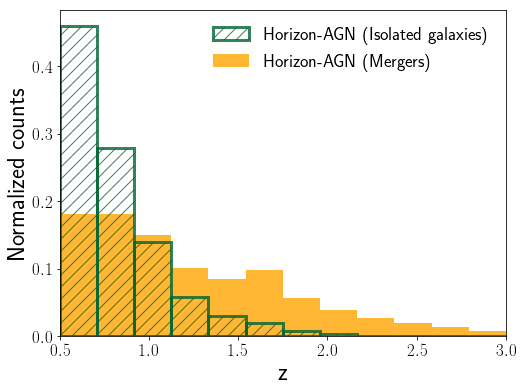}
    \caption{Stellar mass (left) and redshift distribution (right) of the training data (isolated galaxies in striped green and mergers images in solid yellow). We can see obvious deviations between the distributions of isolated galaxies and mergers which should be quantifiable as anomalous by the anomaly WGAN detector.}\label{figure5}
\end{figure*}

Isolated galaxies, on the contrary, satisfy the condition of having only one progenitor when going back in time $1 \textrm{Gyr}$ and only one descendant when going forward in time $1 \textrm{Gyr}$. Images for both datasets are generated as explained in \citep{thesis_Caro}. The final training sample is made of $531,922$ isolated galaxies. Examples of these images are shown in Figure \ref{figure4}. We also show the stellar mass and redshift distributions for our training and test samples in Figure \ref{figure5}. Notice that the distributions are significantly different given the restrictive constraints used to define the sample of isolated galaxies. By imposing no interactions in a $2\ \textrm{Gyr}$ time window we remove very massive galaxies from the sample. This is not a problem since we are aiming to calibrate the sensitivity of the WGAN anomaly detector in identifying out-of-distribution objects. It should be noted that mergers are not anomalous scientifically speaking but are outside of our training data and as such we want to detect them.

\begin{figure*}
    \centering
    \includegraphics[trim={0 0 0 1cm},clip,width=0.9\linewidth]{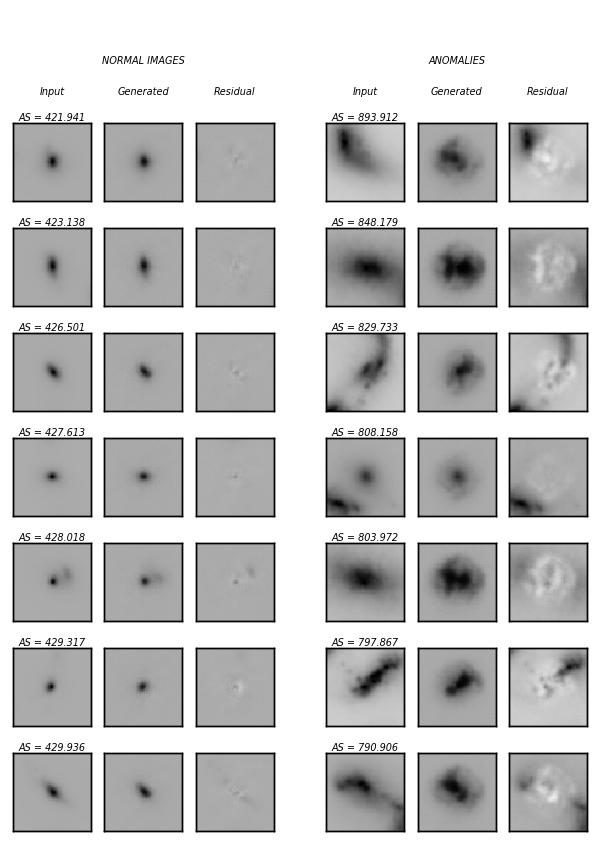}
    \caption{Examples of images draw from the test sample with low anomaly score (left) and and high anomaly score (right). In the first column, we show the input image, in the second column, the closest generated image obtained in the anomaly detection method and in the third column, the residual (pixel by pixel difference) between the input and the generated. The normal images show low anomaly score values and a very similar image can be generated by the network. For the test images the anomaly score is high and no close image can be generated which results in large residuals.}\label{figure6}
\end{figure*}

\begin{figure}
    \centering
    \includegraphics[width=1\linewidth]{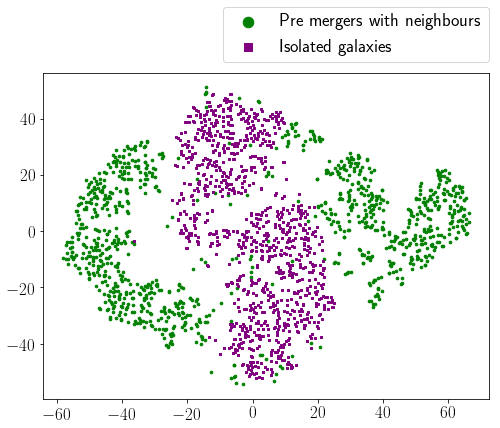}
    \caption{Output of the last convolutional layer of the trained critic after reducing it to 2 dimensions using the t-Distributed Stochastic Neighbor Embedding method (noting that this is just one realisation of the t-SNE). We show where normal images (purple) and images of galaxies with neighbours (green, images that are most different to the normal set) lay in this plane.}\label{figure7}
\end{figure}

We train a WGAN network, described in Section \ref{sec.wgan} for $\sim \times10^6$ epochs using only isolated images. As a first exploratory step, Figure \ref{figure6} shows some examples from the test set indicating their anomaly score, the closest generated image and the residual image (derived by subtracting the closest generated image to the original image). We see that for images with low anomaly score values, the residuals are low as expected, due to the network being able to generate very similar images. For high anomaly score images, the residuals are larger because the network cannot generate a similar image. In several cases, the anomaly is due to the presence of a secondary source as one would naturally expect. Figure \ref{figure7} shows a $2D$ representation of the last layer of the critic (which is used to compute $L_F$) computed with t-Distributed Stochastic Neighbour Embedding \citep[t-SNE,][]{tsne}. t-SNE is a technique for dimensionality reduction that helps with the visualization of high-dimensional datasets. In the high dimensional space, it models the probability distribution that dictates the relationships between neighbours around each point. Then in the low-dimensional space, it recreates, as close as possible, the same distribution. When points are close to one another in the high-dimensional space they will tend to be close to one another in the low-dimensional space as well. We show a subsample of normal galaxies and a subsample of galaxies with a neighbour in the image. This visualization suggests that the network is well trained for our purpose, and the critic is, indeed, able to separate these two classes well. Therefore, using the critic as part of the anomaly score should provide information with which we can detect anomalies. The next sections quantify the performance.

\begin{figure}
    \centering
    \includegraphics[width=1\linewidth]{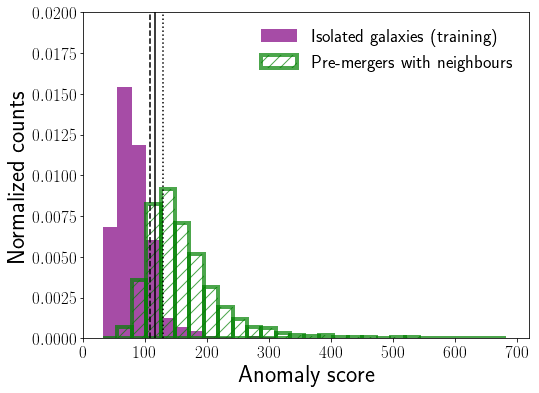}
    \caption{Anomaly score distribution for the isolated galaxies (training data, purple) and the pre-mergers with neighbours (green). The black dash, solid and dotted line represent the three thresholds defined as the value that, respectively, contain $85$, $90$ and $95$ per cent of the training galaxies within the generative distributions. The distribution for the pre-mergers peaks at a higher value than the training data, with clear separations between the distribution. This indicates that the majority of pre-mergers with neighbours are inconsistent with the training sample.}\label{figure8}
\end{figure}

\subsubsection{Results: anomalies caused by a companion in the image}

We now quantify how accurately we can detect anomalies due to the presence of a companion in the image. Given that our training sample consists of isolated galaxies, we can assume that the presence of a secondary object will contribute the most to the degree of anomaly on an image when compared with the training sample.

We, therefore, select all the images with at least a secondary object in the image (this will be our known anomalous sample in this application). All these images belong to the pre-merger phase. However, not all pre-mergers have a secondary object in the image. This is because, in order to speed up the training process and due to memory capability, we have cropped the original images from the simulations to $64\times64$ pixels, and therefore, in some cases, we have artificially removed the companion object that will eventually merge with the main galaxy.

We compute the anomaly score for the training sample as well as for the test sample and compare their distributions in Figure \ref{figure8}. The figure clearly shows that the AS distribution for mergers peaks at larger values than the one for isolated galaxies. To quantify the anomaly detection method, we define a threshold-based method. Images that have an AS larger than the threshold are considered anomalous (or inconsistent with the training sample), while images with AS lower than the threshold, are considered 'normal' (or consistent). We use three different thresholds defined as the value that contains $85$, $90$ and $95$ per cent of the training galaxies within the generative distribution. Using these thresholds, we find $86$, $80$ and $67$ per cent of the anomalous samples are correctly identified as anomalies, respectively. We compare our results with a more traditional method of outlier detection, the k-means clustering method \citep{MacQueen67}. For that, we use non-parametric measures of structure used to quantify the broad morphology: CAS (concentration C, asymmetry A and clumpiness S, \cite{Conselice03}), and Gini/M20 parameters \citep[e.g.][]{Abraham03,Lotz04}. We calculate these parameters for the two samples (training and test) using the code \textsc{statmorph}, a Python package for calculating non-parametric morphological diagnostics of galaxy images \citep{Rodriguez19}, and apply the k-means method with two clusters. We find that $99$ percent of the training sample belongs to one cluster while $74$ per cent of the pre-mergers with a neighbour lie in the other cluster. These results are summarized in Table \ref{table1}.

We further investigate the incorrectly classified pre-merger galaxies, and find that the majority of these are caused by a high flux ratio between the main galaxy and the brightest neighbour. We hence compute the flux ratio $F_r$ between the main galaxy and the companion using \textsc{SExtractor} \citep{Bertin96} and then divide the test sample into three bins depending on the flux ratio ($F_r<1.5$, $1.5<F_r<2$, $F_r>2$). For flux ratios lower than $1.5$, we find that $96$, $93$ and $86$ per cent are correctly classified as anomalous, using the three thresholds mentioned above, respectively. For galaxies between $1.5$ and $2$ times as bright, there is only a small decrease in these percentages ($94$, $90$ and $76$ per cent). It is only when the companion is $2$ times fainter than the central galaxy ($46$ per cent of our test sample) that the percentage of galaxies correctly classified drops to $75$, $66$ and $48$ per cent, respectively, for the three different thresholds. We find that while for $F_r<1.5$ and $1.5<F_r<2$, our results are comparable to the k-means method ($91$ and $93$ per cent of the test sample are correctly classified according to the k-means method), when considering galaxies with a companion $2$ times fainter, the k-means method performs significantly worse at detecting them as anomalous (only $39$ per cent, compared with $75$, $66$ and $48$ per cent, respectively, for the three different thresholds used in our method). These results are summarized in Table \ref{table1}. We have also explored whether the distance between the main galaxy and the companion has an effect on the anomaly score, but we have found no correlation.

\begin{table}
    \caption{Accuracy of the threshold-based anomaly detector for different thresholds, and of k-means anomaly detector method. Each method (columns) is evaluated for test sets (rows) of isolated and pre-merger galaxies. Thresholds $1$, $2$ and $3$ represent the three thresholds defined as the value that contains, respectively, $85$, $90$ and $95$ per cent of the training galaxies within the generative distribution. The last column indicates the accuracy according to the k-means clustering method.}
    \begin{tabular}{c c c c c} 
        \hline
	\multicolumn{5}{c}{Accuracy of the anomaly detector} \\
	\hline
	\hline
        \multicolumn{1}{c}{} & \multicolumn{3}{c|}{Threshold}\\
	& 1 & 2 & 3  & k-means\\ [0.5ex] 
	\hline
	Isolated  & 85\% & 90\% & 95\% & 99\%\\ 
	Pre-mergers & 86\% & 80\% & 67\% & 74\%\\ 
	\hline
    \end{tabular}\label{table1}
\end{table}

\begin{table}
    \caption{Accuracy of the threshold-based anomaly detector for different thresholds, and of k-means anomaly detector method. Each method (columns) is evaluated for test sets (rows) of pre-merger galaxies divided according to their flux ratio between the central galaxy and the brightest neighbour ($F_c/F_n$). Thresholds $1$, $2$ and $3$ represent the three thresholds defined as the value that contains, respectively, $85$, $90$ and $95$ per cent of the training galaxies within the generative distribution. The last column indicates the accuracy according to the k-means clustering method.}
    \begin{tabular}{c c c c c} 
        \hline
        \multicolumn{5}{c}{Accuracy of the anomaly detector} \\
        \hline
        \hline
        \multicolumn{1}{c}{} & \multicolumn{3}{c|}{Threshold}\\
        & 1 & 2 & 3 & k-means\\ [0.5ex] 
        \hline
        $F_r<1.5$   & 96\% & 93\% & 86\% & 91\%\\ [0.5ex] 
        $1.5<F_r<2$ & 94\% & 90\% & 76\% & 93\%\\ [0.5ex] 
        $F_r>2$     & 75\% & 66\% & 48\% & 39\%\\ [0.5ex] 
	\hline
    \end{tabular}\label{table2}
\end{table}

\subsubsection{Results: anomalies caused by merger induced morphological perturbations}

In the previous section, we have seen how the anomaly detection method is able to detect anomalous galaxies with high accuracy when a relatively bright companion is found in the image. Here we investigate how accurately the method works when the companion is not present (i.e. for the pre-merger galaxies that do not show a neighbour object in the image and images of the post-merger phases). This exercise is intended to test the robustness of the WGAN-based anomaly detector given more subtle merger-induced perturbations in the main galaxy light distribution. For this application our anomalous sample, therefore, consists of galaxies in the pre-merger phase that do not have a neighbour object as well as galaxies in the post-merger phases.

\begin{figure}
    \centering
    \includegraphics[width=1\linewidth]{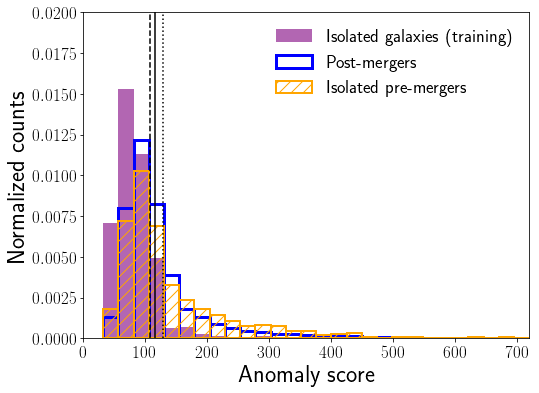}
    \caption{Anomaly score distribution for the isolated galaxies (training data, purple), the isolated pre-mergers (orange) and post-mergers (blue). The black dash, solid and dotted line represent the three thresholds defined as the value that contain, respectively, $85$, $90$ and $95$ per cent of the training galaxies within the generative distribution. The distribution for the isolated pre-mergers and post mergers (test sample) extend to larger values than the training data, even though there is significant overlap between the distribution. There are many samples which are inconsistent with the training data.}\label{figure9}
\end{figure}

Figure \ref{figure9} shows the anomaly score distributions for isolated pre-mergers (or pre-mergers without neighbours) and post-mergers. As expected, we observe that even if both samples have anomaly score distributions extending to larger values than the training set, the majority of the samples overlap with the training data. Using the same three thresholds as before, $46$, $39$ and $30$ per cent of these images are classified correctly (respectively for the three thresholds), while only $8$ per cent when using the k-means clustering method. Overall, it appears the isolated pre-mergers are detected as anomalous slightly more often than the post-mergers. We summarize these results in Table \ref{table3}. These results indicate that the presence of a secondary object is not the only cause of anomaly in these images, although it is the dominating factor, and that the WGAN is able to detect more subtle morphological differences between the samples. Our results also demonstrate the WGAN method is better suited to finding outliers than traditional clustering methods, particularly when the differences with the training sample are subtle and cannot be fully described by global measures, as they are in this case. We further investigate what features describe the difference between the anomalous images and the training set.

\begin{table}
    \addtolength{\tabcolsep}{-2pt} 

    \centering
    \caption{Accuracy of the threshold-based anomaly detector for different thresholds, and of k-means anomaly detector method. Each method (columns) is evaluated for test sets (rows) of isolated pre-mergers, post-merger galaxies and the combinations of both sets. Thresholds $1$, $2$ and $3$ represent the three thresholds defined as the value that contain, respectively, $85$, $90$ and $95$ per cent of the training galaxies within the generative distribution. The last column indicates the accuracy according to the k-means clustering method.}
    \begin{tabular}{c c c c c} 
        \hline
        \multicolumn{5}{c}{Accuracy of the anomaly detector} \\
        \hline
        \hline 
        \multicolumn{1}{c}{} & \multicolumn{3}{c|}{Threshold}\\
        & 1 & 2 & 3 & k-means\\ [0.5ex] 
        \hline
        Isolated pre-mergers+Post-mergers & 46\% & 39\% & 30\% & 8\%\\ [0.5ex] 
        Isolated pre-mergers & 52\% & 45\% & 37\% & 7\%\\ [0.5ex] 
        Post-mergers & 45\% & 38\% & 28\% & 9\%\\ [0.5ex] 
	\hline
    \end{tabular}\label{table3}
\end{table}

\begin{figure}
    \centering
    \includegraphics[width=1\linewidth]{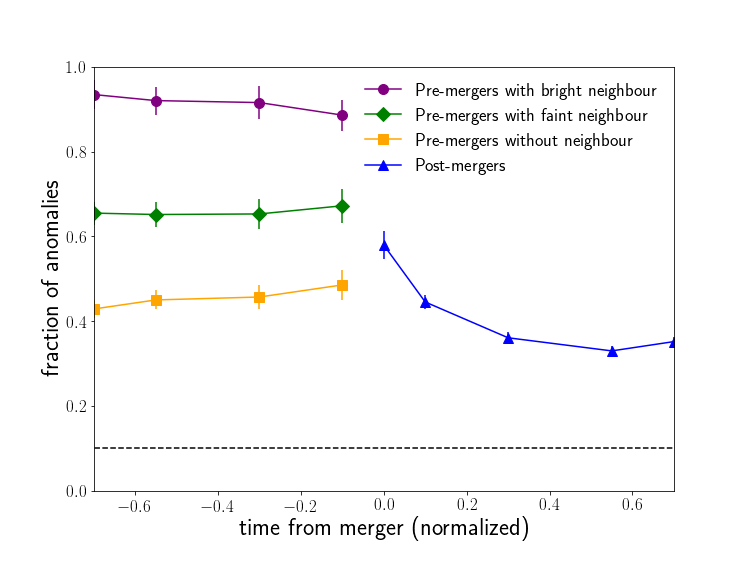}
    \caption{Fraction of samples outside of the training set as a function of time from merger of different types of galaxies. Pre-mergers with a bright neighbour ($F_c/F_n<2$) in purple, pre-mergers with a faint neighbour ($F_c/F_n>2$) in green, isolated pre-mergers in yellow and post-mergers in blue. We show the fraction of samples outside of the training set according to the threshold $2$, which corresponds to $90$ per cent of the training galaxies being consistent.}\label{figure10}
\end{figure}

Figure \ref{figure10} shows the fraction of samples outside of the training set as a function of the position of the image on the merger sequence. The time is normalized such as that $t=-1$ shows the time at which the companion is at $4$ effective radii from the central galaxy, $t=0$ is the time at which the two galaxies become one in the merger tree. We distinguish between pre-mergers with and without neighbours for comparison. Images in which the neighbour is present are the most anomalous, as seen in the previous section. But the fraction does not change significantly with time from the merger. The fraction of anomalous images for the pre-mergers without companions remains mostly constant as well, at around $45$ per cent. In the merger phase, the percentage of samples outside of the training set reaches $56$ per cent and decreases with time from the merger at about $35$ per cent as the system relaxes. The trends we observe prevail even when we use different thresholds, and only differ by a scaling factor.

\begin{figure*}
    \centering
    \includegraphics[width=0.45\linewidth]{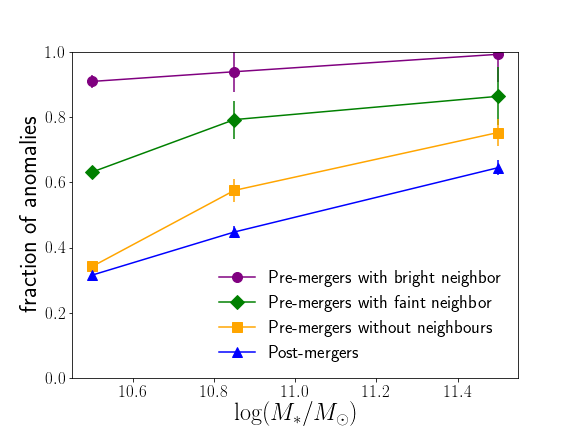}
    \includegraphics[width=0.45\linewidth]{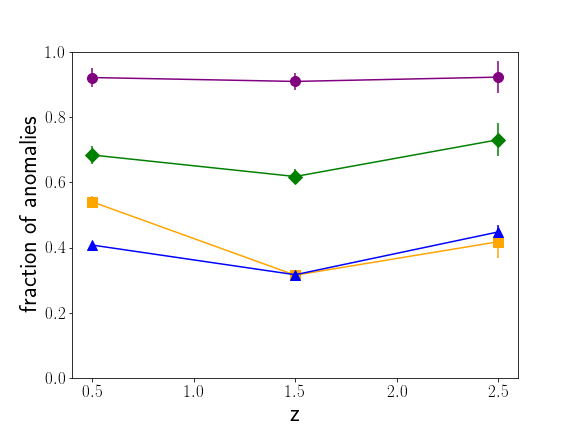}
    \includegraphics[width=0.45\linewidth]{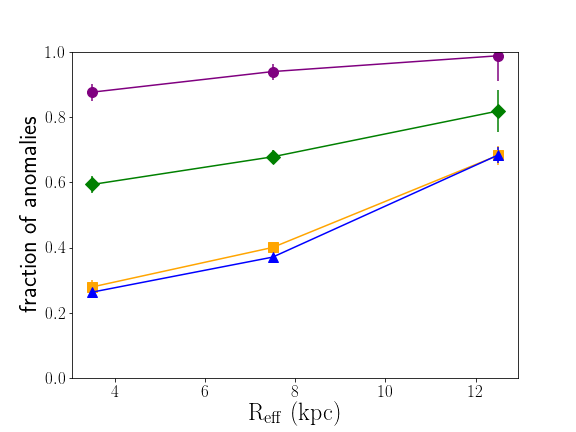}
    \includegraphics[width=0.45\linewidth]{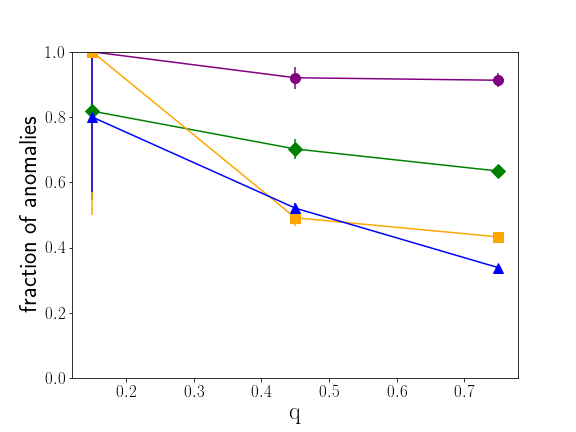}
    \caption{Fraction of samples outside of the training set as a function of stellar mass (top left), redshift $z$ (top right), effective radius $R_eff$ (bottom left) and axis ratio $q$ (bottom right). In blue we show the fraction of post mergers, in yellow the fraction of pre-mergers without a neighbour in the image, in purple the pre-mergers with a bright neighbour and in green pre-mergers with a faint neighbour present in the image. The fractions of anomalies increase with mass and size and decreases with axis ratio, while remains constant with redshift.}\label{figure11}
\end{figure*}

We now investigate other properties from the simulations that can have an effect on the anomaly score: stellar mass, redshift $z$, effective radius $R_{eff}$, and axis ratio $q$. The axis ratio is derived using SExtractor, and the other properties are defined by the simulations. We show in Figure \ref{figure11} how the fraction of samples inconsistent with the training set change as a function of these properties (we again choose the threshold $2$ but note that the choice of the threshold does not affect the trends we observe, they only change by a scaling factor). We observe that there are more anomalous galaxies in the pre-merger stage than in the post-merger phase and that galaxies with a secondary object have, for all properties, higher fractions of anomalies. However, regardless of the galaxy being in the pre- or post-merger phase, or having or not a secondary object, the fraction of anomalies increases with increasing stellar mass and size (larger and more massive galaxies tend to be less consistent with the training set). This is not surprising, as the stellar mass distribution of the training sample does not expand to masses larger than $\log(\textrm{M}_{\ast}/\textrm{M}_{\odot})=10.75$, while for the test sample we find galaxies with stellar mass up to $\log(\textrm{M}_{\ast}/\textrm{M}_{\odot})=12$ (see Figure \ref{figure5}). The anomaly score decreases with axis ratio (the most elongated galaxies are more anomalous). This can be explained by the absence of very elongated galaxies in our training sample. Lastly, we observe that there is not a strong correlation with redshift.

One interesting fact to note here is that the distribution of test images is really being compared to the distribution of training images, and as such, by choosing a hard threshold we quantify rare objects (even in the training set) as being less consistent with the bulk of the rest of training set.

\subsection{Comparison between observations and simulations}

This last application focuses on investigating how we can use the anomaly detection method to compare two sets of data. Here we compare simulated data from Horizon-AGN and data from the CANDELS survey described in Section \ref{sec.data.candels} to see if the WGAN is able to distinguish the images coming from different distributions. Assessing how well modern hydrodynamical simulations reproduce the observed properties of galaxies is a complex task because of the large number of parameters involved. The proposed approach has the advantage of collapsing all properties to one unique metric of similarity that encapsulates all morphological features.

For this application, the first thing we do is to add realistic noise to the mock observations to be able to compare the two datasets directly. We first select sky-only regions from the CANDELS-HST observation in the H-band, to create noise-only mosaics of $64\times 64$ to use randomly with each galaxy from the simulations. For a given galaxy image, we then generate a corresponding Poisson noise, to which we introduced pixel-to-pixel correlation such that 1D auto-correlation power spectral density (PDS) of the sky mosaic matches the source noise. Finally, we add this correlated Poisson noise and a sky mosaic to generate mock images that match the CANDELS-HST observations.

\subsubsection{Training}

\begin{figure*}
 \centering
 \includegraphics[width=0.45\linewidth]{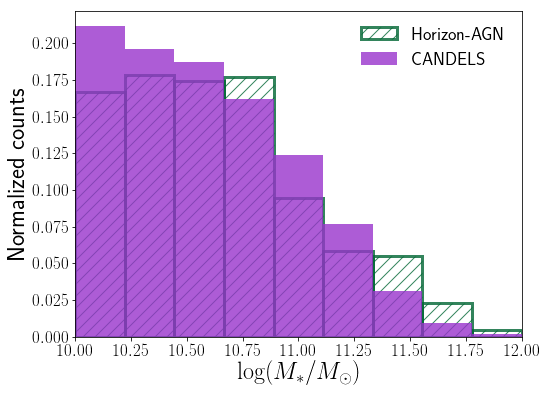}
 \includegraphics[width=0.45\linewidth]{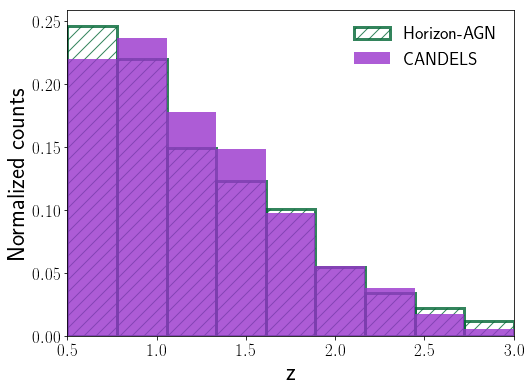}

 \caption{Stellar mass (left) and redshift distribution (right) of the training data (Horizon-AGN images, in striped green), and the test data (CANDELS images, in solid purple).}\label{figure12}
\end{figure*}

We train our WGAN (see Section \ref{sec.wgan}) with a training sample composed of all the mock images from the Horizon-AGN simulation ($1,524,118$ images), with added noise. For this application, the test set is comprised of the $17,611$ CANDELS images in our sample. The stellar mass and redshift distributions for the training (Horizon-AGN) and test set (CANDELS) are shown in Figure \ref{figure12}. Both sets cover the same range in stellar mass and redshift, although with a very slight difference in distribution, so any difference between the datasets is not expected to be a consequence of the selection function.

\begin{figure*}
    \centering
    \includegraphics[trim={0 0 0 1cm},clip,width=0.9\linewidth]{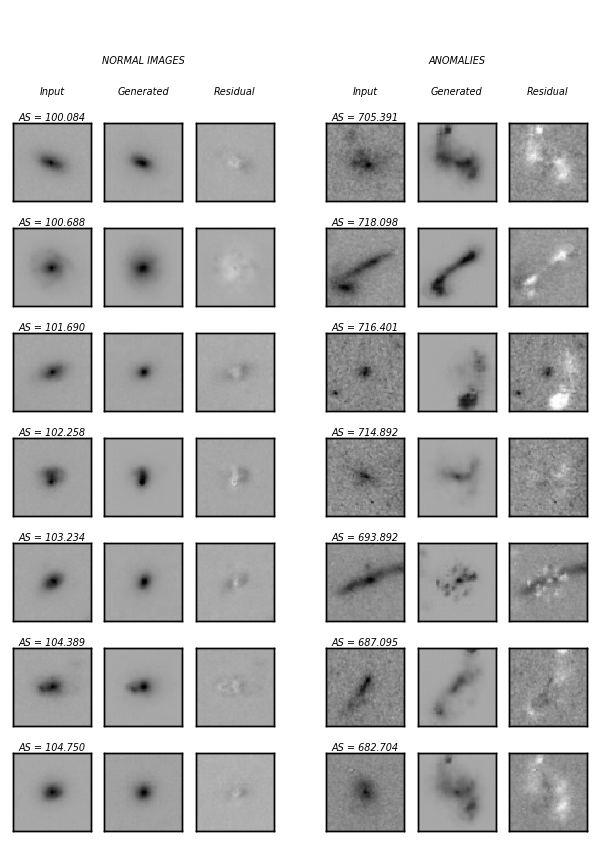}
    \caption{Examples of CANDELS galaxies classified as normal (left) and anomalous (right). Similar to Figure \ref{figure6}, in the first column we show the input image, in the second column, the closest generated image and in the third, the residual image.}\label{figure13}
\end{figure*}

\begin{figure}
    \centering
    \includegraphics[width=1\linewidth]{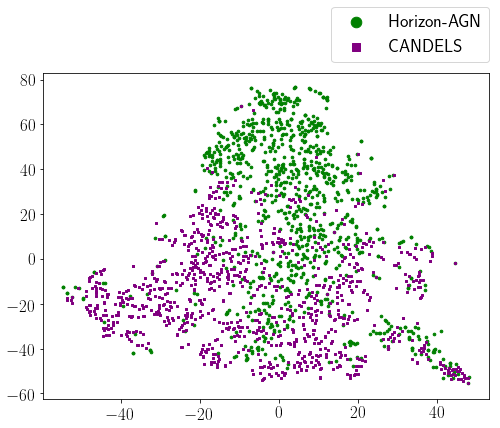}
    \caption{t-SNE space for the output of the last convolutional layer of the trained critic. We show in green a subset of training images and in purple CANDELS images. Note the separation is much less obvious than with Figure \ref{figure7} since the distribution of images is much closer.}\label{figure14}
\end{figure}

Figure \ref{figure13} shows examples of CANDELS galaxies indicating their anomaly score, the closest generated image and the residuals. As in Figure \ref{figure6}, we observe that for galaxies with low anomaly score the network is able to generate a very similar image and, therefore, the residuals are low, while high anomaly scores result in high residuals (the network is not able to generate similar images). Figure \ref{figure14} shows the t-SNE reduced dimensional representation of the output of the last layer of convolutions for both the Horizon-AGN and CANDELS images. We easily observe that the two sets globally populate different parts of the space, although with some degree of overlap, which suggests that the network is able to distinguish the two populations. We investigate the reasons behind this apparent discrepancy in the following section.

\subsubsection{Results: Difference between Horizon-AGN and CANDELS}\label{subsec.candels}

\begin{figure}
    \centering
    \includegraphics[width=1\linewidth]{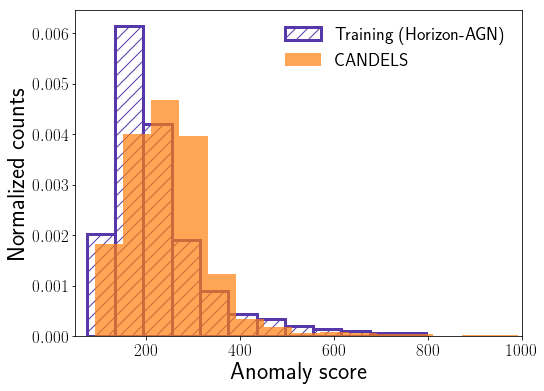}
    \vspace{1ex}
    \caption{Distribution of anomaly scores for the Horizon-AGN galaxies (training sample), in striped blue, and for the CANDELS data in solid orange. The anomaly scores for the CANDELS galaxies are, overall, higher than the training set, which points at differences between the Horizon-AGN simulations and the CANDELS observations.}\label{figure15}
\end{figure}

We compute the anomaly scores for Horizon-AGN and CANDELS images, and show their distribution in Figure \ref{figure15}.

We observe that, even though there is considerable overlap between the two distributions, the anomaly scores for the CANDELS are, overall, higher. If the simulations were to reproduce the observed data perfectly, we would expect the distributions of the anomaly score for both the Horizon-AGN and CANDELS samples to be more consistent. However, the difference in anomaly score distribution suggests that the simulations are not able to completely reproduce the observational data from CANDELS. This could be, in part, because we did not include effects such as dust, but could also include other choices in the physical model of the simulation, or by resolution effects or other prescriptions in the rediative transfer code. Therefore this example has to be seen as an illustration of the potential of this approach to detect global subtle morphological differences between datasets coming from different origins. However, we do not aim to establish robust conclusions given the many limitations. One possible application would be exploring how different simulations, produced with different physical processes, compare with observational data. For that purpose, our WGAN could be trained on images from an observational survey, and then, anomaly scores can be computed for the observational images, and for the images from the different simulations. Comparing the distributions of anomaly scores will give information about which simulations produce images that are more consistent with the observations.

\begin{figure*}
    \centering
    \includegraphics[width=0.45\linewidth]{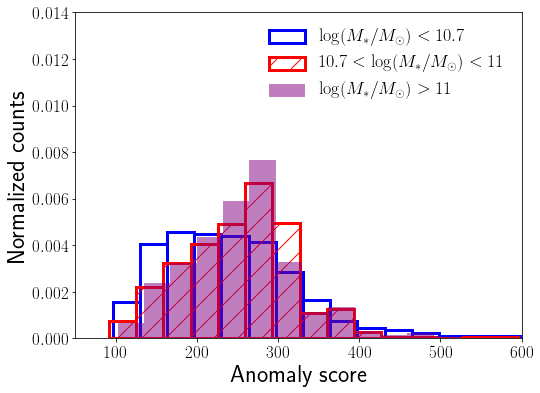}
    \includegraphics[width=0.45\linewidth]{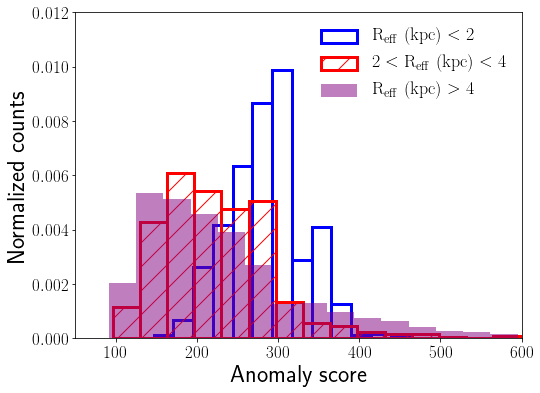}\\
    \includegraphics[width=0.45\linewidth]{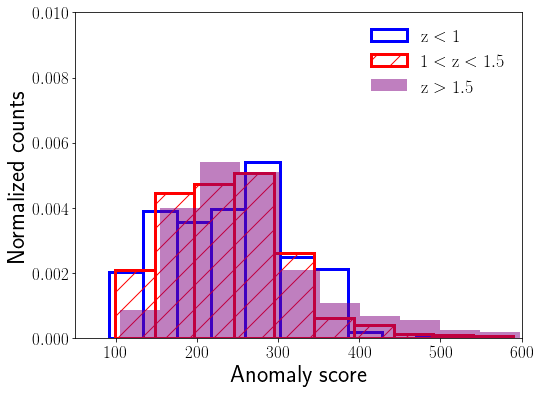}
    \includegraphics[width=0.45\linewidth]{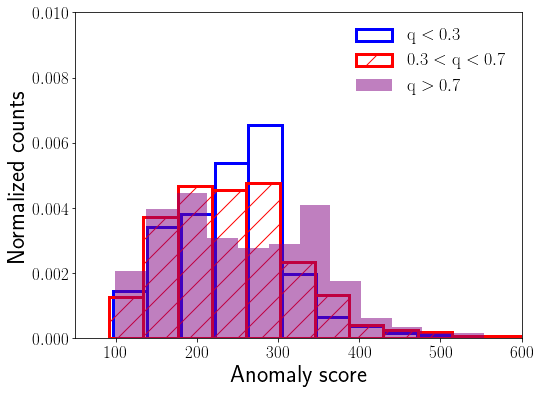}
    \includegraphics[width=0.45\linewidth]{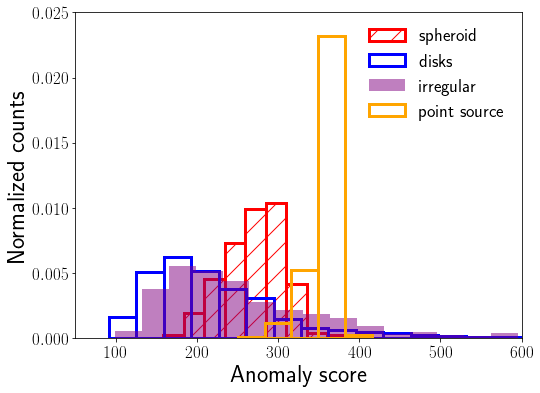}
    \includegraphics[width=0.45\linewidth]{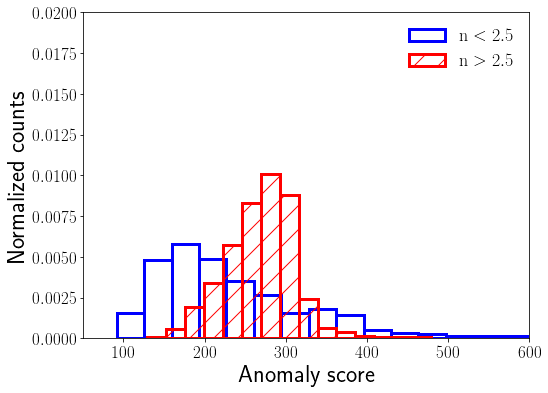}
    \caption{Distribution of the anomaly score for the CANDELS galaxies as a function of different properties (stellar mass, effective radius, redshift, axial ratio, S\'ersic index and morphology type). Each histogram is normalized. These plots show which properties are causing high anomaly scores. We observe that the largest and more massive galaxies tend to have higher anomaly scores, as well as spheroidal galaxies and galaxies with high S\'ersic index.}\label{figure16}
\end{figure*}

As a preliminary step forward, we show in Figure \ref{figure16} the anomaly score distribution for observed galaxies as a function of different galaxy physical properties (stellar mass, effective radius, redshift, axis ratio, S\'ersic index and morphological type). The figure reveals some interesting trends. While there is no effect on the anomaly score due to redshift, globally speaking, massive galaxies tend to be more anomalous. The smallest galaxies tend to have higher AS values, possible due to resolution effects in the simulations. Spheroidal, high S\'ersic index galaxies and point-source/compact are also skewed towards larger AS values. This suggests that compact galaxies might not be well represented in the Horizon-AGN simulation. However, this needs to be investigated further given the limitations of this comparison. 

\section{Summary}\label{sec.summary}

In this first proof-of-concept work, we have explored generative methods as a way to quantify anomalous objects in astronomical imaging data without labels. The method consists of, first, training a WGAN with `normal' data and calculating the anomaly score of the test sample to quantify the degree of anomaly. The main advantage of such an approach is that it can learn complex representations directly from the pixel space without manual extraction of specific features. It can therefore identify subtle morphological differences and collapse morphological comparisons to one unique metric.

We have tested the method on three different applications:
\begin{itemize}
    \item In the first application we asses how accurately we can detect known differences between the training and test samples. In this case, the WGAN is trained with images of isolated galaxies (with no merger history) from the Horizon-AGN simulation, and used to identify known mergers. We show that the WGAN correctly classifies $80$ per cent of the test images as anomalous with a contamination of only $10$ per cent (i.e. $10$ per cent of the isolated galaxies are incorrectly classified as anomalous). The percentage of anomalous images increases to $92$ per cent when not considering images with very faint neighbours in the image. 
    \item In the second application we investigate how our method is able to detect anomalies caused by more subtle properties. When investigating a test sample that consist of images of merging galaxies without a visible companion in the image, we find that $45$ per cent of the test set is anomalous compared to the training set. In this case, the anomaly is cause by morphological features instead of a secondary source in the image. We observe that the most anomalous objects generally have higher mass and have high axis ratio. This is because the training sample lacks galaxies with these properties. It is, therefore, useful to consider how this anomaly detection method really allows us to introspect biases in the training set as well as the physical model with which we can generate such \emph{realistic} training sets.
    \item The third application shows how the anomaly detection method can be used to compare two datasets. The training set for this example is a complete set of simulated galaxies from the Horizon-AGN simulations, and the test or comparison sample comprises observed images from the CANDELS survey. We show that the anomaly score distribution of the observations tends to peak at larger values compared to that of the simulated data. We further explore what properties were causing the main differences, to better understand how the simulations differ from the observations. We observe that the simulations were not reproducing the smallest galaxies and high S\'ersic index galaxies well. This may be, in part, due to the lack of dust treatment in the radiative code, but could also be due to the resolution effects and/or other radiative processes.
\end{itemize}

The code to train our WGAN and generate is made pubic with this work. In future papers we plan to investigate the effect that physical processes from the simulations (such as the addition of dust) have on our analysis when comparing the Horizon-AGN simulation to the CANDELS survey. Additionally, we plan to use the WGAN anomaly detector to look for outliers in the HSC survey (Storey-Fisher et al.) and investigate its applicability in the pipelines of future surveys such as LSST and EUCLID. As part of the efforts to investigate the practical use of generative models to compare simulations to observations, we are also exploring regressive models (Zanisi et al.)

\section*{Acknowledgements}

\bibliographystyle{mnras}
\bibliography{referencias}

\begin{thebibliography}{}
\makeatletter
\relax
\def\mn@urlcharsother{\let\do\@makeother \do\$\do\&\do\#\do\^\do\_\do\%\do\~}
\def\mn@doi{\begingroup\mn@urlcharsother \@ifnextchar [ {\mn@doi@}
  {\mn@doi@[]}}
\def\mn@doi@[#1]#2{\def\@tempa{#1}\ifx\@tempa\@empty \href
  {http://dx.doi.org/#2} {doi:#2}\else \href {http://dx.doi.org/#2} {#1}\fi
  \endgroup}
\def\mn@eprint#1#2{\mn@eprint@#1:#2::\@nil}
\def\mn@eprint@arXiv#1{\href {http://arxiv.org/abs/#1} {{\tt arXiv:#1}}}
\def\mn@eprint@dblp#1{\href {http://dblp.uni-trier.de/rec/bibtex/#1.xml}
  {dblp:#1}}
\def\mn@eprint@#1:#2:#3:#4\@nil{\def\@tempa {#1}\def\@tempb {#2}\def\@tempc
  {#3}\ifx \@tempc \@empty \let \@tempc \@tempb \let \@tempb \@tempa \fi \ifx
  \@tempb \@empty \def\@tempb {arXiv}\fi \@ifundefined
  {mn@eprint@\@tempb}{\@tempb:\@tempc}{\expandafter \expandafter \csname
  mn@eprint@\@tempb\endcsname \expandafter{\@tempc}}}

\bibitem[\protect\citeauthoryear{{Abraham}, {van den Bergh}  \&
  {Nair}}{{Abraham} et~al.}{2003}]{Abraham03}
{Abraham} R.~G.,  {van den Bergh} S.,   {Nair} P.,  2003, \mn@doi [\apj]
  {10.1086/373919}, \href
  {https://ui.adsabs.harvard.edu/abs/2003ApJ...588..218A} {588, 218}

\bibitem[\protect\citeauthoryear{{Abruzzo}, {Narayanan}, {Dav{\'e}}  \&
  {Thompson}}{{Abruzzo} et~al.}{2018}]{Abruzzo18}
{Abruzzo} M.~W.,  {Narayanan} D.,  {Dav{\'e}} R.,   {Thompson} R.,  2018, arXiv
  e-prints, \href {https://ui.adsabs.harvard.edu/abs/2018arXiv180302374A} {p.
  arXiv:1803.02374}

\bibitem[\protect\citeauthoryear{{Arjovsky}, {Chintala}  \&
  {Bottou}}{{Arjovsky} et~al.}{2017}]{Arjovsky17}
{Arjovsky} M.,  {Chintala} S.,   {Bottou} L.,  2017, preprint, \href
  {http://adsabs.harvard.edu/abs/2017arXiv170107875A} {} (\mn@eprint {arXiv}
  {1701.07875})

\bibitem[\protect\citeauthoryear{{Aubert}, {Pichon}  \& {Colombi}}{{Aubert}
  et~al.}{2004}]{Aubert04}
{Aubert} D.,  {Pichon} C.,   {Colombi} S.,  2004, \mn@doi [\mnras]
  {10.1111/j.1365-2966.2004.07883.x}, \href
  {https://ui.adsabs.harvard.edu/abs/2004MNRAS.352..376A} {352, 376}

\bibitem[\protect\citeauthoryear{{Baron} \& {Poznanski}}{{Baron} \&
  {Poznanski}}{2017}]{Baron17}
{Baron} D.,  {Poznanski} D.,  2017, \mn@doi [\mnras] {10.1093/mnras/stw3021},
  \href {https://ui.adsabs.harvard.edu/abs/2017MNRAS.465.4530B} {465, 4530}

\bibitem[\protect\citeauthoryear{{Barro} et~al.,}{{Barro}
  et~al.}{2019}]{Barro19}
{Barro} G.,  et~al., 2019, \mn@doi [\apjs] {10.3847/1538-4365/ab23f2}, \href
  {https://ui.adsabs.harvard.edu/abs/2019ApJS..243...22B} {243, 22}

\bibitem[\protect\citeauthoryear{{Bertin} \& {Arnouts}}{{Bertin} \&
  {Arnouts}}{1996}]{Bertin96}
{Bertin} E.,  {Arnouts} S.,  1996, \mn@doi [A\&AS] {10.1051/aas:1996164}, \href
  {http://adsabs.harvard.edu/abs/1996A%26AS..117..393B} {117, 393}

\bibitem[\protect\citeauthoryear{{Bonjean}, {Aghanim}, {Salom{\'e}}, {Beelen},
  {Douspis}  \& {Soubri{\'e}}}{{Bonjean} et~al.}{2019}]{Bonjean19}
{Bonjean} V.,  {Aghanim} N.,  {Salom{\'e}} P.,  {Beelen} A.,  {Douspis} M.,
  {Soubri{\'e}} E.,  2019, \mn@doi [\aap] {10.1051/0004-6361/201833972}, \href
  {https://ui.adsabs.harvard.edu/abs/2019A&A...622A.137B} {622, A137}

\bibitem[\protect\citeauthoryear{{Boucaud} et~al.,}{{Boucaud}
  et~al.}{2019}]{Boucaud19}
{Boucaud} A.,  et~al., 2019, arXiv e-prints, \href
  {https://ui.adsabs.harvard.edu/abs/2019arXiv190501324B} {p. arXiv:1905.01324}

\bibitem[\protect\citeauthoryear{{Bruzual} \& {Charlot}}{{Bruzual} \&
  {Charlot}}{2003}]{Bruzual03}
{Bruzual} G.,  {Charlot} S.,  2003, \mn@doi [MNRAS]
  {10.1046/j.1365-8711.2003.06897.x}, \href
  {http://adsabs.harvard.edu/abs/2003MNRAS.344.1000B} {344, 1000}

\bibitem[\protect\citeauthoryear{{Cabrera-Vives}, {Reyes}, {F{\"o}rster},
  {Est{\'e}vez}  \& {Maureira}}{{Cabrera-Vives} et~al.}{2017}]{Cabrera17}
{Cabrera-Vives} G.,  {Reyes} I.,  {F{\"o}rster} F.,  {Est{\'e}vez} P.~A.,
  {Maureira} J.-C.,  2017, \mn@doi [\apj] {10.3847/1538-4357/836/1/97}, \href
  {https://ui.adsabs.harvard.edu/abs/2017ApJ...836...97C} {836, 97}

\bibitem[\protect\citeauthoryear{Caro}{Caro}{2018}]{thesis_Caro}
Caro F.,  2018, PhD thesis, l'Observatoire de Paris

\bibitem[\protect\citeauthoryear{{Chabrier}}{{Chabrier}}{2003}]{Chabrier03}
{Chabrier} G.,  2003, \mn@doi [PASP] {10.1086/376392}, \href
  {http://adsabs.harvard.edu/abs/2003PASP..115..763C} {115, 763}

\bibitem[\protect\citeauthoryear{{Conselice}}{{Conselice}}{2003}]{Conselice03}
{Conselice} C.~J.,  2003, \mn@doi [\apjs] {10.1086/375001}, \href
  {http://adsabs.harvard.edu/abs/2003ApJS..147....1C} {147, 1}

\bibitem[\protect\citeauthoryear{{Dahlen} et~al.,}{{Dahlen}
  et~al.}{2013}]{Dahlen13}
{Dahlen} T.,  et~al., 2013, \mn@doi [\apj] {10.1088/0004-637X/775/2/93}, \href
  {http://adsabs.harvard.edu/abs/2013ApJ...775...93D} {775, 93}

\bibitem[\protect\citeauthoryear{{Davidzon} et~al.,}{{Davidzon}
  et~al.}{2019}]{Davidzon19}
{Davidzon} I.,  et~al., 2019, \mn@doi [\mnras] {10.1093/mnras/stz2486}, \href
  {https://ui.adsabs.harvard.edu/abs/2019MNRAS.489.4817D} {489, 4817}

\bibitem[\protect\citeauthoryear{{Dimauro} et~al.,}{{Dimauro}
  et~al.}{2018}]{Dimauro18}
{Dimauro} P.,  et~al., 2018, \mn@doi [\mnras] {10.1093/mnras/sty1379}, \href
  {https://ui.adsabs.harvard.edu/abs/2018MNRAS.478.5410D} {478, 5410}

\bibitem[\protect\citeauthoryear{{Dom{\'\i}nguez S{\'a}nchez},
  {Huertas-Company}, {Bernardi}, {Tuccillo}  \& {Fischer}}{{Dom{\'\i}nguez
  S{\'a}nchez} et~al.}{2018}]{Dominguez18}
{Dom{\'\i}nguez S{\'a}nchez} H.,  {Huertas-Company} M.,  {Bernardi} M.,
  {Tuccillo} D.,   {Fischer} J.~L.,  2018, \mn@doi [\mnras]
  {10.1093/mnras/sty338}, \href
  {https://ui.adsabs.harvard.edu/abs/2018MNRAS.476.3661D} {476, 3661}

\bibitem[\protect\citeauthoryear{{Dubois}, {Devriendt}, {Slyz}  \&
  {Teyssier}}{{Dubois} et~al.}{2012}]{Dubois12}
{Dubois} Y.,  {Devriendt} J.,  {Slyz} A.,   {Teyssier} R.,  2012, \mn@doi
  [\mnras] {10.1111/j.1365-2966.2011.20236.x}, \href
  {https://ui.adsabs.harvard.edu/abs/2012MNRAS.420.2662D} {420, 2662}

\bibitem[\protect\citeauthoryear{{Dubois} et~al.,}{{Dubois}
  et~al.}{2014}]{Dubois14}
{Dubois} Y.,  et~al., 2014, \mn@doi [\mnras] {10.1093/mnras/stu1227}, \href
  {http://adsabs.harvard.edu/abs/2014MNRAS.444.1453D} {444, 1453}

\bibitem[\protect\citeauthoryear{{Frery}, {Habrard}, {Sebban}, {Caelen}  \&
  {He-Guelton}}{{Frery} et~al.}{2017}]{Frery17}
{Frery} J.,  {Habrard} A.,  {Sebban} M.,  {Caelen} O.,   {He-Guelton} L.,
  2017, in {Ceci} M.,  {Hollm{\'e}n} J.,  {Todorovski} L.,  {Vens} C.,
  {D{\v{z}}eroski} S.,  eds, {Machine Learning and Knowledge Discovery in
  Databases}. Springer International Publishing, pp 20--35

\bibitem[\protect\citeauthoryear{Fukushima}{Fukushima}{1988}]{Fukushima88}
Fukushima K.,  1988, Neural Networks, 1, 119

\bibitem[\protect\citeauthoryear{{Fustes}, {Manteiga}, {Dafonte}, {Arcay},
  {Ulla}, {Smith}, {Borrachero}  \& {Sordo}}{{Fustes} et~al.}{2013}]{Fustes13}
{Fustes} D.,  {Manteiga} M.,  {Dafonte} C.,  {Arcay} B.,  {Ulla} A.,  {Smith}
  K.,  {Borrachero} R.,   {Sordo} R.,  2013, \mn@doi [\aap]
  {10.1051/0004-6361/201321445}, \href
  {https://ui.adsabs.harvard.edu/abs/2013A&A...559A...7F} {559, A7}

\bibitem[\protect\citeauthoryear{{Giles} \& {Walkowicz}}{{Giles} \&
  {Walkowicz}}{2018}]{Giles18}
{Giles} D.,  {Walkowicz} L.,  2018, in American Astronomical Society Meeting
  Abstracts \#231. p. 332.03

\bibitem[\protect\citeauthoryear{{Goodfellow}, {Pouget-Abadie}, {Mirza}, {Xu},
  {Warde-Farley}, {Ozair}, {Courville}  \& {Bengio}}{{Goodfellow}
  et~al.}{2014}]{Goodfellow14}
{Goodfellow} I.~J.,  {Pouget-Abadie} J.,  {Mirza} M.,  {Xu} B.,  {Warde-Farley}
  D.,  {Ozair} S.,  {Courville} A.,   {Bengio} Y.,  2014, preprint, \href
  {http://adsabs.harvard.edu/abs/2014arXiv1406.2661G} {} (\mn@eprint {arXiv}
  {1406.2661})

\bibitem[\protect\citeauthoryear{{Grogin} et~al.,}{{Grogin}
  et~al.}{2011}]{Grogin11}
{Grogin} N.~A.,  et~al., 2011, \mn@doi [ApJs] {10.1088/0067-0049/197/2/35},
  \href {http://adsabs.harvard.edu/abs/2011ApJS..197...35G} {197, 35}

\bibitem[\protect\citeauthoryear{{Gulrajani}, {Ahmed}, {Arjovsky}, {Dumoulin}
  \& {Courville}}{{Gulrajani} et~al.}{2017}]{Gulrajani17}
{Gulrajani} I.,  {Ahmed} F.,  {Arjovsky} M.,  {Dumoulin} V.,   {Courville} A.,
  2017, preprint, \href {http://adsabs.harvard.edu/abs/2017arXiv170400028G} {}
  (\mn@eprint {arXiv} {1704.00028})

\bibitem[\protect\citeauthoryear{{Haardt} \& {Madau}}{{Haardt} \&
  {Madau}}{1996}]{Haardt96}
{Haardt} F.,  {Madau} P.,  1996, \mn@doi [\apj] {10.1086/177035}, \href
  {https://ui.adsabs.harvard.edu/abs/1996ApJ...461...20H} {461, 20}

\bibitem[\protect\citeauthoryear{Hassoun}{Hassoun}{1995}]{Hassoun95}
Hassoun M.~H.,  1995, Fundamentals of Artificial Neural Networks, 1st edn.
MIT Press, Cambridge, MA, USA

\bibitem[\protect\citeauthoryear{{Huertas-Company} et~al.,}{{Huertas-Company}
  et~al.}{2015}]{Huertas15}
{Huertas-Company} M.,  et~al., 2015, \mn@doi [\apjs]
  {10.1088/0067-0049/221/1/8}, \href
  {https://ui.adsabs.harvard.edu/abs/2015ApJS..221....8H} {221, 8}

\bibitem[\protect\citeauthoryear{{Huertas-Company} et~al.,}{{Huertas-Company}
  et~al.}{2018}]{Huertas18}
{Huertas-Company} M.,  et~al., 2018, \mn@doi [ApJ] {10.3847/1538-4357/aabfed},
  \href {http://adsabs.harvard.edu/abs/2018ApJ...858..114H} {858, 114}

\bibitem[\protect\citeauthoryear{{Jacobs}, {Glazebrook}, {Collett}, {More}  \&
  {McCarthy}}{{Jacobs} et~al.}{2017}]{Jacobs17}
{Jacobs} C.,  {Glazebrook} K.,  {Collett} T.,  {More} A.,   {McCarthy} C.,
  2017, \mn@doi [\mnras] {10.1093/mnras/stx1492}, \href
  {https://ui.adsabs.harvard.edu/abs/2017MNRAS.471..167J} {471, 167}

\bibitem[\protect\citeauthoryear{{Karras}, {Aila}, {Laine}  \&
  {Lehtinen}}{{Karras} et~al.}{2017}]{Karras17}
{Karras} T.,  {Aila} T.,  {Laine} S.,   {Lehtinen} J.,  2017, preprint, \href
  {http://adsabs.harvard.edu/abs/2017arXiv171010196K} {} (\mn@eprint {arXiv}
  {1710.10196})

\bibitem[\protect\citeauthoryear{{Kaviraj} et~al.,}{{Kaviraj}
  et~al.}{2017}]{Kaviraj17}
{Kaviraj} S.,  et~al., 2017, \mn@doi [\mnras] {10.1093/mnras/stx126}, \href
  {https://ui.adsabs.harvard.edu/abs/2017MNRAS.467.4739K} {467, 4739}

\bibitem[\protect\citeauthoryear{{Kennicutt}}{{Kennicutt}}{1998}]{Kennicutt98}
{Kennicutt} Jr. R.~C.,  1998, \mn@doi [ARA\&A]
  {10.1146/annurev.astro.36.1.189}, \href
  {http://adsabs.harvard.edu/abs/1998ARA%26A..36..189K} {36, 189}

\bibitem[\protect\citeauthoryear{{Kim} \& {Brunner}}{{Kim} \&
  {Brunner}}{2017}]{Edward17}
{Kim} E.~J.,  {Brunner} R.~J.,  2017, \mn@doi [\mnras] {10.1093/mnras/stw2672},
  \href {https://ui.adsabs.harvard.edu/abs/2017MNRAS.464.4463K} {464, 4463}

\bibitem[\protect\citeauthoryear{{Koekemoer} et~al.,}{{Koekemoer}
  et~al.}{2011}]{Koekemoer11}
{Koekemoer} A.~M.,  et~al., 2011, \mn@doi [ApJs] {10.1088/0067-0049/197/2/36},
  \href {http://adsabs.harvard.edu/abs/2011ApJS..197...36K} {197, 36}

\bibitem[\protect\citeauthoryear{{Komatsu} et~al.,}{{Komatsu}
  et~al.}{2011}]{Komatsu11}
{Komatsu} E.,  et~al., 2011, \mn@doi [\apjs] {10.1088/0067-0049/192/2/18},
  \href {https://ui.adsabs.harvard.edu/abs/2011ApJS..192...18K} {192, 18}

\bibitem[\protect\citeauthoryear{{Laigle} et~al.,}{{Laigle}
  et~al.}{2019}]{Laigle19}
{Laigle} C.,  et~al., 2019, \mn@doi [\mnras] {10.1093/mnras/stz1054}, \href
  {https://ui.adsabs.harvard.edu/abs/2019MNRAS.486.5104L} {486, 5104}

\bibitem[\protect\citeauthoryear{{Lotz}, {Primack}  \& {Madau}}{{Lotz}
  et~al.}{2004}]{Lotz04}
{Lotz} J.~M.,  {Primack} J.,   {Madau} P.,  2004, \mn@doi [\aj]
  {10.1086/421849}, \href
  {https://ui.adsabs.harvard.edu/abs/2004AJ....128..163L} {128, 163}

\bibitem[\protect\citeauthoryear{MacQueen}{MacQueen}{1967}]{MacQueen67}
MacQueen J.~B.,  1967, in Cam L. M.~L.,  Neyman J.,  eds,  Vol. 1, Proc. of the
  fifth Berkeley Symposium on Mathematical Statistics and Probability.
  University of California Press, pp 281--297

\bibitem[\protect\citeauthoryear{{Meusinger}, {Schalldach}, {Scholz}, {in der
  Au}, {Newholm}, {de Hoon}  \& {Kaminsky}}{{Meusinger}
  et~al.}{2012}]{Meusinger12}
{Meusinger} H.,  {Schalldach} P.,  {Scholz} R.~D.,  {in der Au} A.,  {Newholm}
  M.,  {de Hoon} A.,   {Kaminsky} B.,  2012, \mn@doi [\aap]
  {10.1051/0004-6361/201118143}, \href
  {https://ui.adsabs.harvard.edu/abs/2012A&A...541A..77M} {541, A77}

\bibitem[\protect\citeauthoryear{{Murphy}, {Linden}, {Dong}, {Hensley},
  {Momjian}, {Helou}  \& {Evans}}{{Murphy} et~al.}{2018}]{Murphy18}
{Murphy} E.~J.,  {Linden} S.~T.,  {Dong} D.,  {Hensley} B.~S.,  {Momjian} E.,
  {Helou} G.,   {Evans} A.~S.,  2018, \mn@doi [ApJ] {10.3847/1538-4357/aac5f5},
  \href {http://adsabs.harvard.edu/abs/2018ApJ...862...20M} {862, 20}

\bibitem[\protect\citeauthoryear{Nash}{Nash}{1950}]{Nash50}
Nash J.~F.,  1950, \mn@doi [Proceedings of the National Academy of Sciences]
  {10.1073/pnas.36.1.48}, 36, 48

\bibitem[\protect\citeauthoryear{{Nayyeri} et~al.,}{{Nayyeri}
  et~al.}{2017}]{Nayyeri17}
{Nayyeri} H.,  et~al., 2017, \mn@doi [\apjs] {10.3847/1538-4365/228/1/7}, \href
  {http://adsabs.harvard.edu/abs/2017ApJS..228....7N} {228, 7}

\bibitem[\protect\citeauthoryear{Neyshabur, Bhojanapalli  \&
  Chakrabarti}{Neyshabur et~al.}{2017}]{Neyshabur17}
Neyshabur B.,  Bhojanapalli S.,   Chakrabarti A.,  2017, CoRR, abs/1705.07831

\bibitem[\protect\citeauthoryear{{Pasquet}, {Bertin}, {Treyer}, {Arnouts}  \&
  {Fouchez}}{{Pasquet} et~al.}{2019}]{Pasquet19}
{Pasquet} J.,  {Bertin} E.,  {Treyer} M.,  {Arnouts} S.,   {Fouchez} D.,  2019,
  \mn@doi [\aap] {10.1051/0004-6361/201833617}, \href
  {https://ui.adsabs.harvard.edu/abs/2019A&A...621A..26P} {621, A26}

\bibitem[\protect\citeauthoryear{{Protopapas}, {Giammarco}, {Faccioli},
  {Struble}, {Dave}  \& {Alcock}}{{Protopapas} et~al.}{2006}]{Protopapas06}
{Protopapas} P.,  {Giammarco} J.~M.,  {Faccioli} L.,  {Struble} M.~F.,  {Dave}
  R.,   {Alcock} C.,  2006, \mn@doi [\mnras]
  {10.1111/j.1365-2966.2006.10327.x}, \href
  {https://ui.adsabs.harvard.edu/abs/2006MNRAS.369..677P} {369, 677}

\bibitem[\protect\citeauthoryear{{Radford}, {Metz}  \& {Chintala}}{{Radford}
  et~al.}{2015}]{Radford15}
{Radford} A.,  {Metz} L.,   {Chintala} S.,  2015, preprint, \href
  {http://adsabs.harvard.edu/abs/2015arXiv151106434R} {} (\mn@eprint {arXiv}
  {1511.06434})

\bibitem[\protect\citeauthoryear{{Ravanbakhsh}, {Lanusse}, {Mandelbaum},
  {Schneider}  \& {Poczos}}{{Ravanbakhsh} et~al.}{2016}]{Ravanbakhsh16}
{Ravanbakhsh} S.,  {Lanusse} F.,  {Mandelbaum} R.,  {Schneider} J.,   {Poczos}
  B.,  2016, arXiv e-prints, \href
  {https://ui.adsabs.harvard.edu/abs/2016arXiv160905796R} {p. arXiv:1609.05796}

\bibitem[\protect\citeauthoryear{{Rodriguez-Gomez} et~al.,}{{Rodriguez-Gomez}
  et~al.}{2015}]{Rodriguez-Gomez15}
{Rodriguez-Gomez} V.,  et~al., 2015, \mn@doi [\mnras] {10.1093/mnras/stv264},
  \href {https://ui.adsabs.harvard.edu/abs/2015MNRAS.449...49R} {449, 49}

\bibitem[\protect\citeauthoryear{{Rodriguez-Gomez} et~al.,}{{Rodriguez-Gomez}
  et~al.}{2019}]{Rodriguez19}
{Rodriguez-Gomez} V.,  et~al., 2019, \mn@doi [\mnras] {10.1093/mnras/sty3345},
  \href {https://ui.adsabs.harvard.edu/abs/2019MNRAS.483.4140R} {483, 4140}

\bibitem[\protect\citeauthoryear{Salimans, Goodfellow, Zaremba, Cheung,
  Radford, Chen  \& Chen}{Salimans et~al.}{2016}]{Salimans16}
Salimans T.,  Goodfellow I.,  Zaremba W.,  Cheung V.,  Radford A.,  Chen X.,
  Chen X.,  2016, in Lee D.~D.,  Sugiyama M.,  Luxburg U.~V.,  Guyon I.,
  Garnett R.,  eds, , Advances in Neural Information Processing Systems 29.
Curran Associates, Inc., pp 2234--2242, \url
  {http://papers.nips.cc/paper/6125-improved-techniques-for-training-gans.pdf}

\bibitem[\protect\citeauthoryear{{Salpeter}}{{Salpeter}}{1955}]{Salpeter55}
{Salpeter} E.~E.,  1955, \mn@doi [\apj] {10.1086/145971}, \href
  {https://ui.adsabs.harvard.edu/abs/1955ApJ...121..161S} {121, 161}

\bibitem[\protect\citeauthoryear{{Santini} et~al.,}{{Santini}
  et~al.}{2014}]{Santini14}
{Santini} P.,  et~al., 2014, \mn@doi [\aap] {10.1051/0004-6361/201322835},
  \href {http://adsabs.harvard.edu/abs/2014A%26A...562A..30S} {562, A30}

\bibitem[\protect\citeauthoryear{{Schlegl}, {Seeb{\"o}ck}, {Waldstein},
  {Schmidt-Erfurth}  \& {Langs}}{{Schlegl} et~al.}{2017}]{Schlegl17}
{Schlegl} T.,  {Seeb{\"o}ck} P.,  {Waldstein} S.~M.,  {Schmidt-Erfurth} U.,
  {Langs} G.,  2017, preprint, \href
  {http://adsabs.harvard.edu/abs/2017arXiv170305921S} {} (\mn@eprint {arXiv}
  {1703.05921})

\bibitem[\protect\citeauthoryear{{Solarz}, {Bilicki}, {Gromadzki}, {Pollo},
  {Durkalec}  \& {Wypych}}{{Solarz} et~al.}{2017}]{Solarz17}
{Solarz} A.,  {Bilicki} M.,  {Gromadzki} M.,  {Pollo} A.,  {Durkalec} A.,
  {Wypych} M.,  2017, \mn@doi [\aap] {10.1051/0004-6361/201730968}, \href
  {https://ui.adsabs.harvard.edu/abs/2017A&A...606A..39S} {606, A39}

\bibitem[\protect\citeauthoryear{{Sreejith} et~al.,}{{Sreejith}
  et~al.}{2018}]{Sreejith18}
{Sreejith} S.,  et~al., 2018, \mn@doi [\mnras] {10.1093/mnras/stx2976}, \href
  {https://ui.adsabs.harvard.edu/abs/2018MNRAS.474.5232S} {474, 5232}

\bibitem[\protect\citeauthoryear{{Stefanon} et~al.,}{{Stefanon}
  et~al.}{2017}]{Stefanon17}
{Stefanon} M.,  et~al., 2017, \mn@doi [\apjs] {10.3847/1538-4365/aa66cb}, \href
  {https://ui.adsabs.harvard.edu/abs/2017ApJS..229...32S} {229, 32}

\bibitem[\protect\citeauthoryear{{Sutherland} \& {Dopita}}{{Sutherland} \&
  {Dopita}}{1993}]{Sutherland93}
{Sutherland} R.~S.,  {Dopita} M.~A.,  1993, \mn@doi [\apjs] {10.1086/191823},
  \href {https://ui.adsabs.harvard.edu/abs/1993ApJS...88..253S} {88, 253}

\bibitem[\protect\citeauthoryear{{Teyssier}}{{Teyssier}}{2002}]{Teyssier02}
{Teyssier} R.,  2002, \mn@doi [\aap] {10.1051/0004-6361:20011817}, \href
  {http://adsabs.harvard.edu/abs/2002A%26A...385..337T} {385, 337}

\bibitem[\protect\citeauthoryear{Thanh{-}Tung, Tran  \& Venkatesh}{Thanh{-}Tung
  et~al.}{2019}]{Hoang19}
Thanh{-}Tung H.,  Tran T.,   Venkatesh S.,  2019, CoRR, abs/1902.03984

\bibitem[\protect\citeauthoryear{{Tuccillo}, {Huertas-Company},
  {Decenci{\`e}re}, {Velasco-Forero}, {Dom{\'{\i}}nguez S{\'a}nchez}  \&
  {Dimauro}}{{Tuccillo} et~al.}{2018}]{Tuccillo18}
{Tuccillo} D.,  {Huertas-Company} M.,  {Decenci{\`e}re} E.,  {Velasco-Forero}
  S.,  {Dom{\'{\i}}nguez S{\'a}nchez} H.,   {Dimauro} P.,  2018, \mn@doi
  [MNRAS] {10.1093/mnras/stx3186}, \href
  {http://adsabs.harvard.edu/abs/2018MNRAS.475..894T} {475, 894}

\bibitem[\protect\citeauthoryear{{Tweed}, {Devriendt}, {Blaizot}, {Colombi}  \&
  {Slyz}}{{Tweed} et~al.}{2009}]{Tweed09}
{Tweed} D.,  {Devriendt} J.,  {Blaizot} J.,  {Colombi} S.,   {Slyz} A.,  2009,
  \mn@doi [\aap] {10.1051/0004-6361/200911787}, \href
  {https://ui.adsabs.harvard.edu/abs/2009A&A...506..647T} {506, 647}

\bibitem[\protect\citeauthoryear{{Wong}, {Moore}, {Cooper}  \& {Wagner}}{{Wong}
  et~al.}{2005}]{Wong03}
{Wong} W.-K.,  {Moore} A.,  {Cooper} G.,   {Wagner} M.,  2005, Journal of
  Machine Learning Research, 6, 1961

\bibitem[\protect\citeauthoryear{{Zhang}, {Vukotic}  \& {Gardner}}{{Zhang}
  et~al.}{2018}]{Zhang18}
{Zhang} J.,  {Vukotic} I.,   {Gardner} R.,  2018, preprint (\mn@eprint {arXiv}
  {1801.10094})

\bibitem[\protect\citeauthoryear{van~der Maaten \& Hinton}{van~der Maaten \&
  Hinton}{2008}]{tsne}
van~der Maaten L.,  Hinton G.,  2008, Journal of Machine Learning Research, 9,
  2579

\makeatother
\end{thebibliography}
\clearpage

\appendix
\section{Training algorithm} \label{appendix}

Given a batch of real and generated images, the critic is trained for $n_\textrm{critic}$ iterations to approximate the Wasserstein distance, by maximising the loss in equation \eqref{eq.loss} whilst keeping the weights of the generator fixed. Afterwards, the generator's weight are updated for a single iteration, by maximising equation \eqref{eq.gen_loss}, whilst the critic weights are held constant so that it minimizes the approximate Wasserstein distance. 
This is repeated until the network has converged. Figure \ref{figure2} shows a schematic representation the WGAN training procedure. 

\begin{algorithm}
\SetAlgoLined
\While{$\boldsymbol{\theta}$ has not converged}{
    \For{$t=1,...,n_{critic}$}{
        \For{$i = 1,..., m$}{
            Sample $\bold{x} \sim P_r$, real data.\;
            Sample $\bold{z} \sim P_z$, latent variable.\;
            Sample $\epsilon \sim U[0,1]$, random number to apply gradient penalty.\;
            $\tilde{\boldsymbol{x}} \leftarrow G_{\theta}(\boldsymbol{z})$, generate image from latent variable\;
            $\hat{\boldsymbol{x}} \leftarrow \epsilon \boldsymbol{x} + (1-\epsilon)\tilde{\boldsymbol{x}}$, apply gradient penalty\;
            $L^{(i)} \leftarrow C_\psi(\tilde{\boldsymbol{x}}) - C_\psi(\boldsymbol{x}) + \lambda(\|\nabla_{\hat{\boldsymbol{x}}}C_\psi(\hat{\boldsymbol{x}})\|_2 - 1)^2$, calculate loss.}
        $\boldsymbol{\psi} \leftarrow Adam(\nabla_\psi\sum^m_{i=1} L^{(i)}/m, \psi, \alpha, \beta_1, \beta_2)$\;}
        \For{$i = 1,...,m$}{
            Sample a batch of latent variables ${\boldsymbol{z}^{(i)}} \sim P_z$.\;}
        $\boldsymbol{\theta} \leftarrow Adam(\nabla_{\theta}\sum^m_{i=1} -C_\psi(G_{\theta}(\boldsymbol{z}^{(i)}))/m, w, \alpha, \beta_1, \beta_2)$\;
        }
 \caption{WGAN with gradient penalty training algorithm \protect\citep{Gulrajani17}. We use default values gradient penalty coefficient of $\lambda = 10$, number of critic iterations per generator iteration $n_\textrm{critic} = 10$, batch size  $m = 64$ and Adam hyperparameters: $\alpha =0.00005$, $\beta_1=0.5$, $\beta_2=0.9$.}
\end{algorithm}

\end{document}